\documentclass[%
 reprint,
 amsmath,amssymb,
 aps,
floatfix,
]{revtex4-1}

\usepackage{graphicx}  
\usepackage{dcolumn}   
\usepackage{bm}        
\usepackage{verbatim}  
\usepackage{mathtools}
\usepackage{ulem}
\usepackage{preamble}
\usepackage[utf8]{inputenc}
\usepackage[usenames,dvipsnames]{color}
\usepackage{soul}
\usepackage{bm}
\usepackage{xcolor}
\usepackage{subfigure}
\usepackage{longtable}
\DeclareUnicodeCharacter{2212}{\textendash{}}
\DeclareUnicodeCharacter{0338}{\o}
\graphicspath{{figs/}}

\makeatletter
\newcommand{\printfnsymbol}[1]{%
  \textsuperscript{\@fnsymbol{#1}}%
}
\makeatother

\begin{document}
\title{ Theory of coupled ion-electron transfer kinetics } 

\author{Dimitrios Fraggedakis$^1$}
\thanks{email: dimfraged@gmail.com}
\author{Michael McEldrew$^1$}
\author{Raymond B. Smith$^1$}
\author{Yamini Krishnan$^1$}
\author{Yirui Zhang$^{2}$}
\author{Peng Bai$^{1,3}$}
\author{William C. Chueh$^4$}
\author{Yang Shao-Horn$^{2,5}$}
\author{Martin Z. Bazant$^{1,6}$}
\thanks{email: bazant@mit.edu}

\affiliation{
   $^1$Department of Chemical Engineering, Massachusetts Institute of Technology, Cambridge, MA 02139 USA
   }
\affiliation{
   $^2$Departments of Mechanical Engineering, Massachusetts Institute of Technology, Cambridge, MA 02139 USA
   }
\affiliation{
  $^3$Department of Energy, Environmental and Chemical Engineering, Washington University, Saint Louis, MO 63130 USA
   }
\affiliation{
  $^4$Department of Materials Science and Engineering, Stanford University, Stanford, CA 94305, USA
   }
 \affiliation{
   $^5$Department of Materials Science and Engineering, Massachusetts Institute of Technology, Cambridge, MA 02139 USA
   }
\affiliation{
  $^6$Department of Mathematics, Massachusetts Institute of Technology,  Cambridge, MA 02139 USA
   }

\date{\today}

\begin{abstract}
The microscopic theory of chemical reactions is based on transition state theory, where atoms or ions transfer classically over an energy barrier, as electrons maintain their ground state. Electron transfer is fundamentally different and occurs by tunneling in response to solvent fluctuations. Here, we develop the theory of coupled ion-electron transfer, in which ions and solvent molecules fluctuate cooperatively to facilitate {non-adiabatic} electron transfer. We derive a general formula of the reaction rate that depends on the overpotential, solvent properties, the electronic structure of the electron donor/acceptor, and the excess chemical potential of ions in the transition state. For Faradaic reactions, the theory predicts curved Tafel plots with a concentration-dependent reaction-limited current. For moderate overpotentials, our formula reduces to the Butler-Volmer equation and explains its relevance, not only in the well-known limit of large electron-transfer (solvent reorganization) energy, but also in the opposite limit of large ion-transfer energy.  The rate formula is applied to Li-ion batteries, where reduction of the electrode host material couples with ion insertion. In the case of lithium iron phosphate, the theory accurately predicts the concentration dependence of the exchange current measured by {\it in operando} X-Ray microscopy without any adjustable parameters. These results pave the way for interfacial engineering to enhance ion intercalation rates, not only for batteries, but also for ionic separations and neuromorphic computing.
\end{abstract}

\maketitle

\section{Introduction}
\label{sec:intro}

\begin{figure*}[!ht]
\centering
\includegraphics[width=1\textwidth]{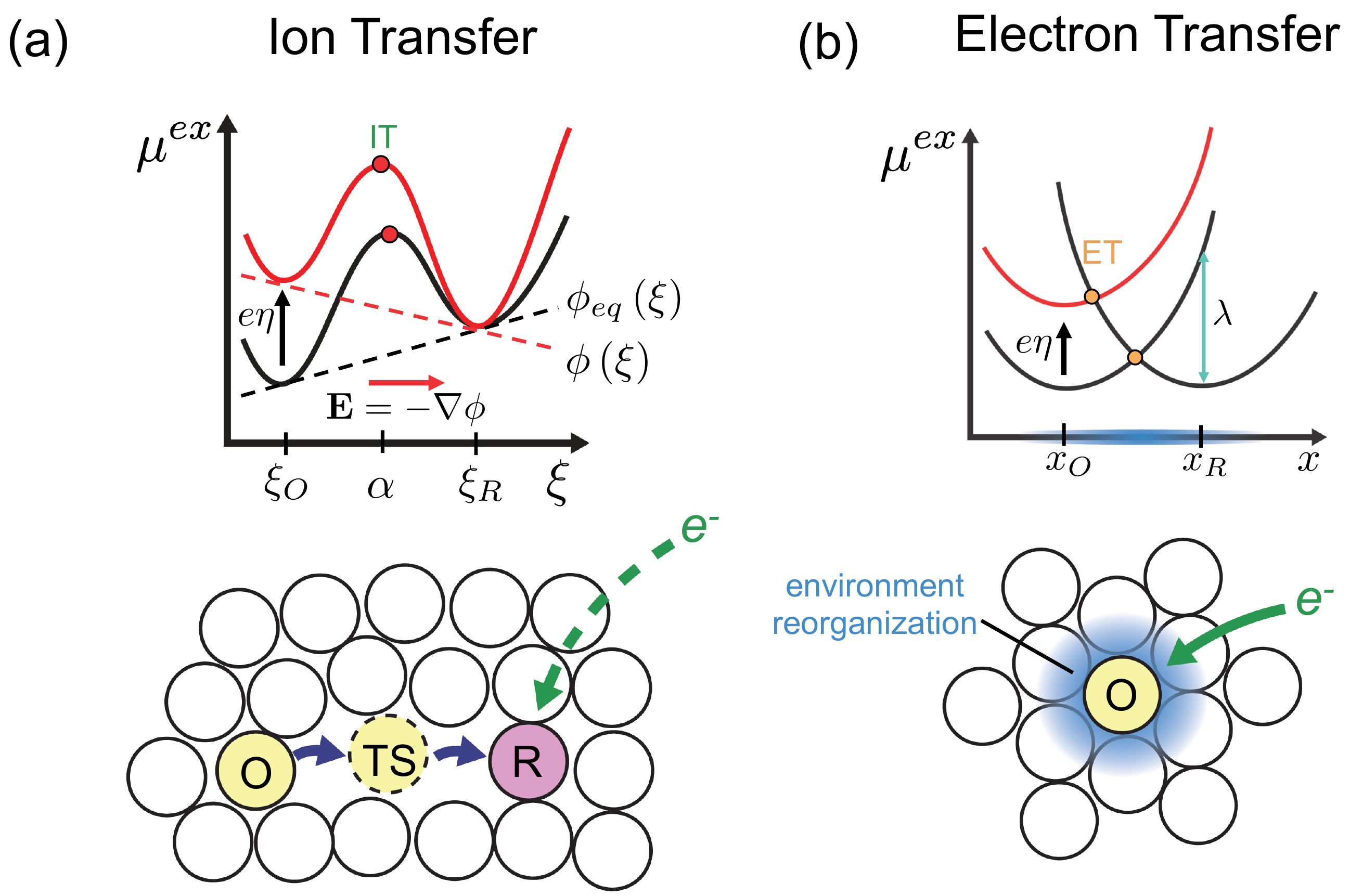}\hspace{8pt}
\caption{ Physical pictures of ion (a) and electron (b) transfer, in terms of landscapes of the excess chemical potential $\mu_{ex}$. In both cases, $O$, TS and $R$ correspond to the oxidized, transition, and reduced states, respectively. (a) For ion transfer, the reaction coordinate $\xi$ is the the distance of the ion from its reduces state, e.g. the electrode where the charge transfer takes place. Under an applied overpotential $\eta$, a cation, for example, is attracted towards the surface of the electrode, where at some distance (red point) towards the electrode it is reduced by an electron (dashed green arrow). This distance is described by the charge transfer coefficient $\alpha$. (b) For electron transfer, the reaction coordinate $x$ corresponds to the environment polarization coordinate. For an electron transfer to occur, both the reactant and product states need to be at the same energy (orange point) where the electron is able to tunnel between the two states. The reorganization step, which is related to the reorganization energy $\lambda$ of the environment, is denoted using the blue shade.}
\label{fig:review_mech}
\end{figure*}

Charge transfer reactions are paramount in biology, e.g. in protein-protein electron transfer~\cite{nocek1996theory,antonyuk2013structures,jeuken2003conformational} and photosynthesis~\cite{bixon1995kinetic,wang2007protein}, and electrochemical engineering, e.g. in water desalination~\cite{dykstra2017theory,zhang2018faradaic,he2016faradaic,singh2018theory,singh2019timeline} and energy conversion and storage~\cite{newman2004,bazant2013,lim2016origin,winter2018before,xu2014electrolytes}. Well established models, such as Butler-Volmer (BV) and Marcus kinetics, are available to describe the rate of charge transfer carried by ions or electrons~\cite{kuznetsov_book,SchmicklerText,bard2001,bazant2013}, respectively, as sketched in Fig.~\ref{fig:review_mech}. Here, we consider concerted or {\it coupled} ion-electron transfer reactions, which have received much less attention and lack a simple rate formula. 

Efforts to describe charge-transfer reactions can be traced to the early twentieth century. Building on Tafel's discovery of exponential overpotential dependence for Faradaic reaction rates~\cite{tafel1905polarisation}, the seminal work of Butler~\cite{butler1924part3} and Volmer~\cite{erdey-gruz1931zur} introduced the phenomenological BV equation, which is by far the most widely used rate expression in electrochemistry~\cite{bard2001,SchmicklerText,bockris1998modern} and electrochemical engineering~\cite{newman2004,biesheuvel2009imposed}. The classical derivation of the BV equation~\cite{bockris1998modern} is based on Eyring's transition-state theory~\cite{eyring1935activated} applied to ion transfer (IT), Fig.~\ref{fig:review_mech}(a)~\cite{biesheuvel2009imposed}. The IT transition state is assumed to be fixed at a distance $\alpha d$ from the reduced state, where $d$ is the distance from the oxidized state. Electron transfer (ET) is not treated explicitly in this picture, but the IT rate expression is sometimes interpreted to imply that a fraction of electrons $\alpha$ is transferred on the reduced side of the transition-state barrier, while the remaining fraction $1-\alpha$ is transferred on the oxidized side in order to complete the reaction~\cite{SchmicklerText,norskov2014fundamental}. The BV equation has recently been generalized for consistency with non-ideal thermodynamics~\cite{bazant2013}, e.g. for phase transformations driven by charge transfer reactions~\cite{bazant2017thermodynamic}, setting the stage for our analysis below. 

Marcus was the first to recognize that classical transition state theory cannot be applied to electrons~\cite{marcus1956electrostatic}. Instead, he proposed that ET occurs iso-energetically in response to the environment fluctuations of the RedOx species, whose energy landscape is modeled as intersecting parabolas in terms of the solvent reorganization  coordinate~\cite{marcus1964chemical,marcus1956electrostatic,marcus1956oxidationreduction,marcus1957oxidationreduction,marcus1959electrochemical,marcus1960exchange,marcus1963oxidationreduction}, Fig.~\ref{fig:review_mech}(b). In the case of outer sphere ET reactions, Marcus related the ET activation energy barrier to the solvent reorganization energy $\lambda$, which is dominated by the dielectric polarization of the medium. Soon afterwards, Hush derived a similar rate expression for adiabatic inner sphere ET reactions, where the electrons transfer in response to molecular vibrations, and thus the reorganization energy is dominated by the phonons of the molecular complex~\cite{hush1957electrode,hush1958adiabatic,hush1961adiabatic,hush1968homogeneous}. Notably, the Marcus-Hush rate expression reduces to the BV equation for overpotentials much smaller than the reorganization energy~\cite{bard2001}.  

The quantum mechanical theory of ET, pioneered by Levich, Dogonadze, Chizmadzhev, Christov, and Kuznetsov~\cite{dogonadze1962kinetics,levich1963osnovnie,dogonadze1965theory,christov2012collision,christov1975quantum,kuznetsov_book}, also leads to Marcus-Hush rate expressions. In general, electron transfer depends on the interaction of the electrons that participate in the reaction with the environment of the particular RedOx active sites~\cite{marcus1993,kuznetsov_book}. When fluctuations of the environment are large enough to make the reduced and oxidized states iso-energetic, ET occurs by tunneling with a probability controlled by electronic coupling between the RedOx states. In the case of Faradaic ET reactions, all available electrons can participate in the reaction, so the Marcus-Hush rate must be integrated over the Fermi distribution of electron energies in the band structure~\cite{marcus1965electron}. Thus, for the typical case of a metallic electrode~\cite{chidsey1991,henstridge2012marcus,zeng2014simple}, the energy landscape of the electron donor is described by a family of parabolas, Fig.~\ref{fig:review_mech}(b).

Classical theories assume either IT or ET is the rate limiting step, but there are situations where both processes occur simultaneously. The characteristic example is coupled proton-electron transfer (CPET)~\cite{Koper2013,Saveant2014,Mayer2004,Hammes-Schiffer2018,Saveant2008,Fukuzumi2012}. The theory of CPET is based on the ideas of electron transfer, where the fluctuating environment of the RedOx species determines how the reaction will proceed. In addition to the isoenergetic requirement for the ET part, the initial and final vibrational state of the proton bond with the molecular complex need also to be at the same energy before the proton transfer occurs~\cite{cukier1998proton,reece2009proton,Hammes-Schiffer2018}. Once the isoenergetic conditions for both the ET and PT are satisfied, both the electron and the proton are transferred through tunneling. Although the mathematical framework and concepts behind CPET are mature~\cite{horvath2012insights,hammes2008proton,melander2020grand} and validated several times~\cite{parada2019concerted}, there has still been little attention paid to the limit where the {\it ions behave classically} and only the electrons are quantum particles~\cite{bazant2013,Schmickler1996}, as in the case of ion intercalation~\cite{bai2014}. 

Ion intercalation has been traditionally modeled by the BV equation in the context of batteries~\cite{newman2004,doyle1993} without any mention to electron transfer. Additionally, ion intercalation is used for selective separations and water desalination~\cite{porada2017nickel,singh2018theory,zhang2018faradaic,singh2019timeline}, where the process is thought to be purely classical and dominated by IT. However, it has been recently shown~\cite{bai2014} that electron transfer can be the limiting step in ion intercalation in the context of Li-ion batteries. Therefore, a complete microscopic picture of ion intercalation is still lacking, as both IT and ET seem to be important during the process. 

In this work, we develop the fundamental theory of coupled ion-electron transfer (CIET) reactions, in which ions and electrons are transferred through a concerted mechanism. Starting from the framework of far-from equilibrium chemical thermodynamics~\cite{keizer2012statistical} applied to charge-transfer reactions~\cite{bazant2013}, we derive a simple, closed-form reaction-rate formula, which takes into account: (i) the non-ideal thermodynamics of the reactants and products, (ii) ionic configurational entropy and other non-idealities in the transition state, (iii) the electrostatic coupling between the ions and the electrons, (iv) the tunneling of electrons, (v) the solvation effects of the ions near interfaces, and (vi) the electronic density of states and quantum statistics of the electron donor. Interestingly, our formula reduces to the BV equation in two different limits of moderate overpotentials, when either ET or IT is dominant. In the case of ion intercalation coupled to a metallic electron donor, our formula reduces to Marcus-Hush-Chidsey (MHC) kinetics of ET with a new pre-factor accounting for the crowding of ions during IT. The theory accurately predicts the exchange current versus concentration for LiFePO$_4$ (LFP) obtained directly from x-ray imaging experiments~\cite{lim2016origin}, as well as the chronoamperometry data~\cite{bai2014}, and paves the way for predictive  modeling of Li-ion battery reaction kinetics~\cite{smith2017multiphase}.

\section{Physical Picture}

\begin{figure*}[!ht]
\centering
\includegraphics[width=1\textwidth]{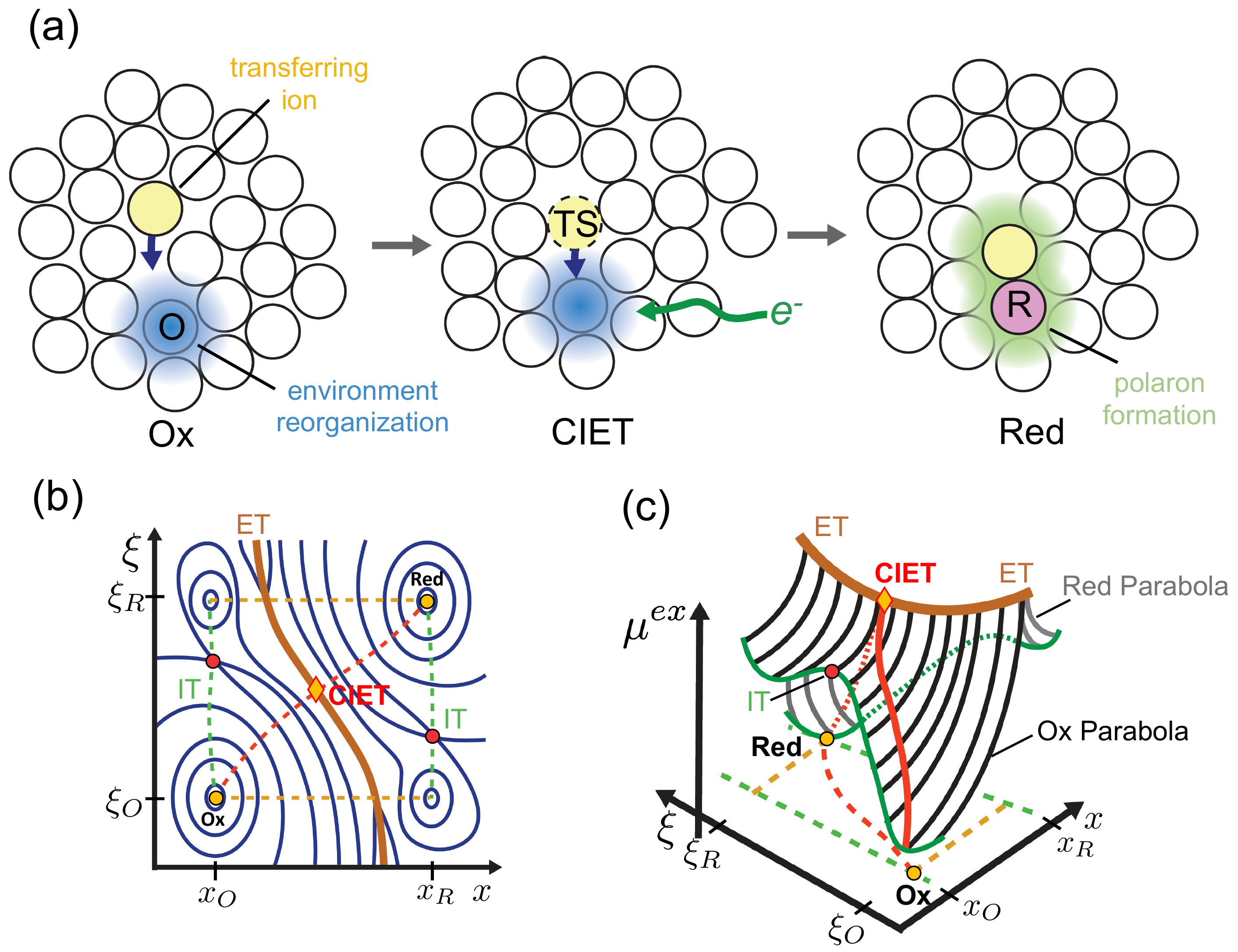}\hspace{8pt}
\caption{Physical and energy picture of coupled ion-electron transfer. (a) Schematic representation of an amorphous medium that consists of neutral particles, reduced, and oxidized species. At first, the transferring ion starts moving towards the site that is going to be reduced (blue shade), where its environment becomes reorganized as a result of thermal fluctuations. Once the environment of the reactant and product states have similar energies and also when the ion is at the transition state (TS), an electron coming either from a metal or a dopant will tunnel and reduce the site. Concurrently with this event, an ion-polaron pair is formed. (b) \& (c) Energy landscape drawn as a contour plot and a three-dimensional surface in terms of the ionic $\xi$ and polarization $x$ coordinates. There are four minima, two of which correspond to the RedOx states. These are accessed only through a single point (yellow diamond) that has the lowest energy barrier and allows the transfer of both ions and electrons at their product state. In the contour plot, the minimum energy path is depicted with the red dashed line, while in the three-dimensional surface with the solid and thinly-dashed red lines. Additionally, in the three-dimensional landscape the solid and thinly-dashed green are the intersection of the energy landscape with $x$-normal planes at $x=x_O$ and $x=x_R$, and finally, the thick-dashed red is the projection of the minimum energy path on $\mu^{ex}=0$.}
\label{fig:general_ciet}
\end{figure*}

In order to develop a general physical picture of coupled ion-electron transfer, we consider a medium consisting of neutral particles, unpaired cations and electroneutral cation-anion pairs (polarons), Fig~\ref{fig:general_ciet}(a). The concerted ion-electron transfer requires both the ions and the electrons that participate in the reaction to be transferred together to form the product complex. In the present picture, the electron transfer part corresponds to the reduction of a neutral site by either delocalized (coming from a metal) or localized (from dopants or impurities) electrons, and the ion transfer corresponds to the physical transfer of the cation nearby the reduced site. We assume that the individual completion of the steps (ion or electron transfer) cannot take place due to the prohibitively large electrostatic energy required to separate the cation and the reduced site from their final state. 

The physical picture of Fig.~\ref{fig:general_ciet}(a) can be translated into the energy landscape shown in Figs.~\ref{fig:general_ciet}(b) \& (c). Our representation of coupled ion-electron transfer is a combination of the classical ion~\cite{bazant2013} and electron~\cite{marcus1985electron} transfer treatments as described in Figs.~\ref{fig:review_mech}. More specifically, the ionic coordinate is equivalent to the distance travelled by the ion to reach its final state $\xi$, and the electron coordinate is the solvent polarization one $x$. Given the non-adiabatic nature of electron transfer, the oxidized and reduced states are described by different parabolic function for constant values of $\xi$, Figs.~\ref{fig:general_ciet}(b) \& (c). 

We consider the energy landscape to have the four local minima as clearly shown in the contour plot of Figs.~\ref{fig:general_ciet}(b). These minima represent the following cases: i) at $\left(x,\xi\right)=\left(x_O,\xi_O\right)$ both ions and electrons are in reactant/oxidized state, ii) at $\left(x,\xi\right)=\left(x_R,\xi_R\right)$ both ions and electrons are in product/reduced state, iii) at $\left(x,\xi\right)=\left(x_O,\xi_R\right)$ the ion transfer is completed, but the electron transfer has not occurred yet, iv) at $\left(x,\xi\right)=\left(x_R,\xi_O\right)$ the electron has tunneled in the reduced state, while the ion has not moved from its initial position. The system can explore the last two minima only when the driving force (overpotential) is large enough to exceed the electrostatic attraction between the ion and the reduced complex in their final state.

According to Figs.~\ref{fig:general_ciet}(b) \& (c), the non-adiabatic surfaces intersect each other along the dark orange line. Across the intersection, electrons can tunnel from the reactant to the product state as their environments are at the same energy state. However, only one point along the electron transfer line corresponds to the minimum energy barrier (yellow diamond - CIET), where both the ion and the electron transfers occur at the same time.

Similar ideas to coupled ion-electron transfer have been previously demonstrated for electrocatalytic adsorption reactions, where solvated ions transfer at the electrode interface where an electron transfer occurs and covalent bonding takes place~\cite{Schmickler1995,Schmickler1996,Koper1998,Santos2008,Santos2009}. Another example is that of non-adsorbing RedOx reactions near the electrodes. In that case, the ions have to work against the formed double layer to reach the electrified interface, where along their way an electron is transferred to the ion which consequently moves back to the solution~\cite{Hartnig2003,Saveant2008,lin2016electrical,Limaye2020}. In both examples, the concerted nature of the process translates into a multidimensional energy landscape in the reaction coordinates~\cite{sumi1986dynamical,Hartnig2003}, similar to that shown in Figs.~\ref{fig:general_ciet}(b) \& (c).

\section{Theory}
\label{sec:model}
\subsection{Thermodynamics of RedOx Reactions}
We consider a general electrochemical reaction of the form
$$
    O^{+} + {e}^- \leftrightarrows
    R
$$
where $O^+$ and $R$ represent oxidized and reduced states, respectively, which may involve multiple ions or neutral molecules, while ${e}^-$ corresponds to the electron which participates in the RedOx reaction. 
In the general theory of electrochemical thermodynamics~\cite{bazant2013}, the electrochemical potential of individual species is described in terms of its diffusional chemical potential $\mu_i$, which is defined relative to a reference state $\Theta$, as a function of the electrical potential, $\phi$, and species activity, $a_i$
\begin{align}
    \mu_i = \frac{\delta G}{\delta c_i} = \mu_i^\Theta + k_{B}T\ln{a_i} + z_{i}e\phi 
    = \mu_i^{ex} + k_{B}T\ln{c_i}
    \label{}
\end{align}
where $c_i$ dimensionless species concentrations, $k_{B}$ is the Boltzmann constant, $T$ is the absolute temperature, $e$ is the elementary charge, and $G$ corresponds to the non-equilibrium free energy of the system that can also be defined in terms of reaction coordinates. The excess chemical potential is defined as
\begin{equation}
    \mu_i^{ex}=\mu_i^\Theta+k_BT\ln\gamma_i+z_ie\phi    
\end{equation}
where $\gamma_i=a_i/c_i$ is the activity coefficient of species $i$ and contains all the present non-idealities of the studied system at its reduced and oxidized states (e.g. chemical or mechanical effects).

\subsection{Reaction Kinetics}\label{sub:Rxn_s}
Coupled ion-electron transfer reactions take place at interfacial regions in thermodynamically non-ideal systems and involve electron tunneling events. For this reason, CIET reactions require a description of reacting species that accounts for thermodynamic non-idealities, the transition state and the tunneling process. We build the theory using the Keizer's principles of nonequilibrium statistical mechanics~\cite{keizer2012statistical}, as formulated for electrochemical reactions in~\cite{bazant2013}.

The reaction rate is written in terms of elementary processes as
\begin{align}
    R_{r,o} = R_{red} - R_{ox}
\label{eq:R_rxn_P}
\end{align}
where $R_{red}$ and $R_{ox}$ correspond to the reduction and oxidation reaction rates, respectively~\cite{bazant2013}. Each of these rates is further analyzed as separate probabilistic events leading to
\begin{subequations}
\begin{align}
    R_{red} \sim \textrm{P}\left(O\right)\textrm{P}\left(O\rightarrow O^{\ddag}\right) \textrm{P}\left(ET|O^{\ddag}\right)
\label{eq:R_rxn_fwd}\\
    R_{ox} \sim \textrm{P}\left(R\right)\textrm{P}\left(R\rightarrow R^{\ddag}\right) \textrm{P}\left(ET|R^{\ddag}\right)
\label{eq:R_rxn_rev}
\end{align}
\end{subequations}
In the above two equations, $\textrm{P}\left(O/R\right)$ corresponds to the probability on finding particles of the oxidized or the reduced species as well as electrons and holes from an electron donor {and is proportional to the species concentration $n_{e/h}c_{O/R}e^{-w_{O/R}/k_BT}$. The part $c_{O/R}e^{-w_{O/R}/k_BT}$ corresponds to the RedOx species concentration at the reaction site, where $w_{O/R}$ represents the free energy required to form the RedOx species from a chemical reservoir~\cite{bazant2013,smith2017multiphase,fedorov2014} and can be associated to the electric double layer and/or species adsorption energies at the electrode/electrolyte interface~\cite{trefalt2016charge}}. Also, $n_{e}$ and $n_h$ are the normalized concentrations of the electrons and holes~\cite{kuznetsov_book}, respectively. $\textrm{P}\left(O/R\rightarrow O^{\ddag}/R^{\ddag}\right)$ describes the probability of thermally exciting the oxidized/reduced species to a state at which electron tunneling becomes iso-energetic, and is proportional to the Boltzmann factor relative to the transition state and local equilibrium excess chemical potential $e^{\left(\mu^{ex}_\ddag-\mu^{ex}_{O/R}\right)/k_BT}$. Also, $\textrm{P}\left(ET|O^{\ddag}/R^{\ddag}\right)\equiv k_T$ corresponds to the conditional probability of a successful electron tunneling event~\cite{kuznetsov_book}, given that the oxidized/reduced species are thermally activated. Thus, the formal expressions for the forward and backward rates read~\cite{bazant2013}
\begin{subequations}
\begin{align}
    R_{red,\varepsilon} = k_0k_T n_e c_O\exp\left(-{\frac{w_O}{k_BT}}\right) \exp\left(-{\frac{\mu^{ex}_\ddag-\mu^{ex}_{O}}{k_BT}}\right)
\label{eq:R_rxn_fwd_full}\\
    R_{ox,\varepsilon} = k_0k_T n_h c_R \exp\left(-{\frac{w_R}{k_BT}}\right) \exp\left(-{\frac{\mu^{ex}_\ddag-\mu^{ex}_{R}}{k_BT}}\right)
\label{eq:R_rxn_rev_full}
\end{align}
\label{eq:R_rxn_fwd_bck_full}
\end{subequations}
{\color{blue}where $k_0$ is a prefactor that satisfies microscopic reversibility~\cite{kondepudi2014modern,sekimoto2010} and $\mu_O^{ex}$ includes also the excess chemical potential of the electrons $\mu_e^{ex}$. Depending on the context of the reaction one is interested in (e.g. electron transfer in bulk solution or near an electrode), $k_0$ has different physical meaning. In bulk electron transfer reactions under dilute conditions, for instance, $k_0$ corresponds to the attempt frequency per unit volume~\cite{kuznetsov_book}, while for electron transfer reaction near electrodes it corresponds to the the attempt frequency per unit area. In more complex reactions, where all microscopic processes are clearly described (e.g. adsorption of ions on electrode surfaces), $k_0$ can have an Arrhenius form. } Last, the hole concentration can be expressed as $n_h=1-n_e$~\cite{kuznetsov_book,SchmicklerText}.

\subsection{Coupled ion-electron transfer}\label{sec:TST_s}

In the present section, we derive the model for $\mu^{ex}_\ddag$ found in eqs.~\ref{eq:R_rxn_fwd_bck_full} for the case of coupled ion-electron transfer. Electron transfer reactions have been modeled successfully using classical Marcus theory~\cite{marcus1993}. In general, both $\mu_O^{ex}$ and $\mu_R^{ex}$ can be extended to include the reaction coordinate dependencies. We propose a description of the transition state that includes the typical harmonic polarization reaction coordinate $x$~\cite{bazant2013}, as well as an additional term that accounts for the ion transfer
\begin{subequations}
\begin{align}
    \mu_{O}^{ex}(x,\xi) = \mu_{O}^{ex}\left(x_O,\xi_O\right) + \frac{\kappa_O}{2}\left(x-x_O\right)^2 + f_O\left(x,\xi\right)
\label{eq:mu_O_mrc} \\
    \mu_{R}^{ex}(x,\xi) = \mu_{R}^{ex}\left(x_R,\xi_R\right) + \frac{\kappa_R}{2}\left(x-x_R\right)^2 + f_R\left(x,\xi\right)
\label{eq:mu_R_mrc}
\end{align}
\label{eq:mu_OR_mrc}
\end{subequations}
The functions $f_O$ and $f_R$ describe dependencies with respect to an additional reaction coordinate, the ionic one $\xi$, which accounts for non-idealities arising from the ion transfer reaction, as well as its coupling to the electron transfer. Here $\xi$ takes the values $\xi_O$ in the oxidized state and $\xi_R$ in the reduced state and can be interpreted as the distance the ion has to move for the ion transfer to happen. That implies the following conditions on $f_O$ and $f_R$
\begin{align}
    f_O\left(x_O,\xi_O\right) = f_R\left(x_R,\xi_R\right) = 0
\label{eq:f_O_xo_f_R_xr_eq}
\end{align}
which ensure that reactant/product complexes satisfy their equilibrium thermodynamics description and 
$$\mu_{O}(x_{O},\xi_{O})=\mu_{O}^\Theta + k_{B}T\ln{c_{O}\gamma_{O}} + z_{O}e\phi(x_{O},\xi_{O}),$$
$$\mu_{R}(x_{R},\xi_{R})=\mu_{R}^\Theta + k_{B}T\ln{c_{R}\gamma_{R}} + z_{R}e\phi(x_{R},\xi_{R})$$

In the classical electron transfer theory, where the reaction does not depend on the ionic coordinate, reduction of the oxidized species occurs iso-energetically~\cite{kuznetsov_book}. Therefore, both the reactant and product environments need to have exactly the same energy for an electron to be transferred - electron tunneling events must conserve energy. This is true when $x=x_\ddag$, where 
\begin{align}
\mu_{O}^{ex}\left(x_\ddag\right)=\mu_{R}^{ex}\left(x_\ddag\right)=\mu_{\ddag,ET}^{ex}\left(x_\ddag\right).
\label{eq:mu_O_R_ddag}
\end{align}
is used to determine $x_\ddag$~\cite{bazant2013}. Here, $\mu_{\ddag,ET}^{ex}$ is the TS chemical potential defined at the intersection of the parabolas. The value of $x_\ddag$ from eq.~\ref{eq:mu_O_R_ddag} results in the same activation barrier as the quantum mechanical approach of ET using Fermi's golden rule~\cite{kuznetsov_book,zwanzig2001nonequilibrium}. 

\begin{figure*}[!ht]
    \centering
    \includegraphics[width=1\textwidth]{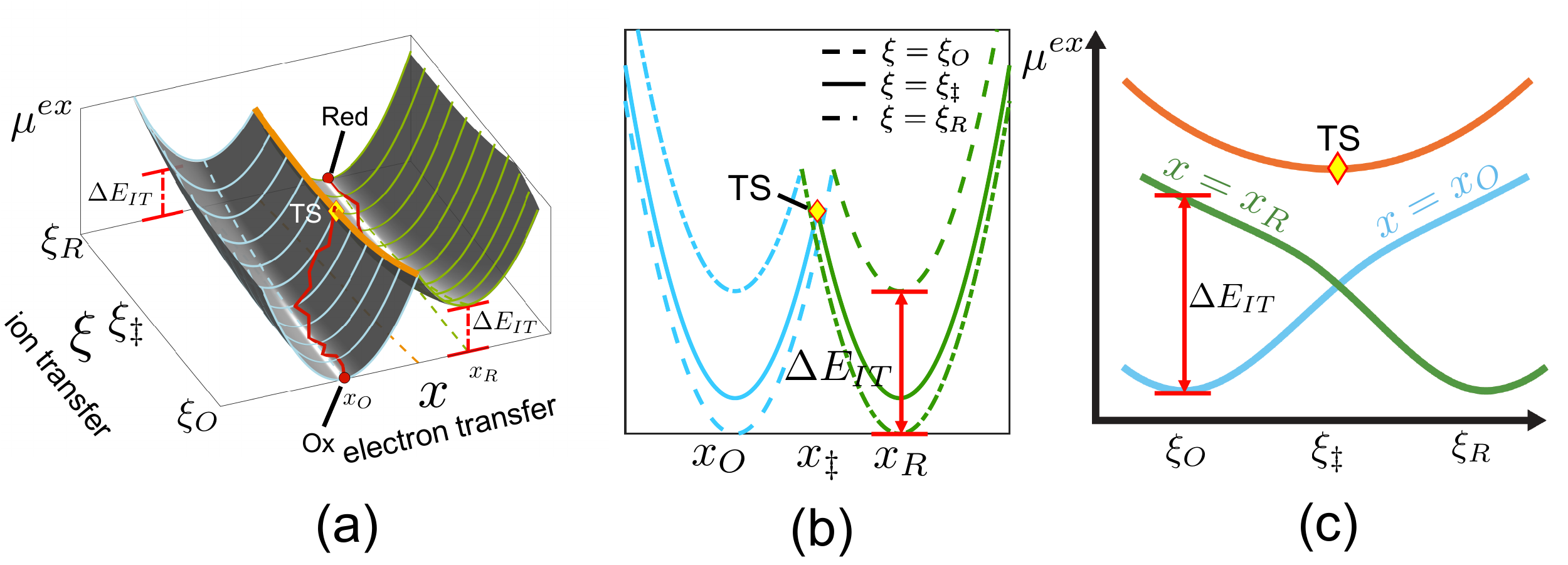}\hspace{8pt}
    \caption{ (a) Excess chemical potential landscape for both reactants and products vs. the reaction coordinates $\left(x,\xi\right)$. The orange line corresponds to the common points of the two parabolas, where iso-energetic electron transfer is possible. The red line depicts a fluctuating path in the two dimensional space where it passes through the maximum with the minimum value of the intersecting parabolas (yellow diamond), enabling the transfer of both the ion and the electron. In this picture, the coupled ion-electron transfer corresponds to concerted reaction process. (b) Excess chemical potential landscape for $\xi=\xi_O$ (dashed line), $\xi=\xi_\ddag$ (solid line), $\xi=\xi_R$ (dashed-dot line). (c) Schematic illustration of the projection of the energy profiles of the oxidized ($x=x_O$) and reduced ($x=x_R$) species on the energy-$\xi$ plane is shown.}
    \label{fig:lndscpe}
\end{figure*} 

Figure~\ref{fig:lndscpe} illustrates the energy landscape including the ionic reaction coordinate introduced in eqs.~\ref{eq:mu_O_mrc} \& \ref{eq:mu_R_mrc}. The idea of including the ionic coordinates for the reaction landscape can be seen as a generalization of Marcus's original picture~\cite{marcus1964chemical}. The consideration of the ionic coordinate allows for the description of environmental effects on the transition state barrier, e.g. site exclusion due to surface crowding phenomena~\cite{bazant2013,bazant2017thermodynamic}. This is in contrast to existing studies that focus on dilute liquids and solids.

Fig.~\ref{fig:lndscpe}(a) shows the intersection of the 2D parabolas of both reactant and product species (orange thick line). Solving eq.~\ref{eq:mu_OR_mrc} at the intersection, we can express $x_\ddag$ in terms of $\xi$. In principle, ET is possible for any value of $\xi$ along the orange line, and thus the electron or ion transfer occurs separately from each other. However, we expect the ion-electron transfer to be concerted, as opposed to sequential, due to the large electrostatic attraction between the electron and the ion at their product state - the ion and the electron reside nearby each other in ion intercalation materials, where they interact through short-range electrostatics. For example, in the case where $\xi=0$ there is a probability for electron transfer without ion transfer. The energy barrier for that process, though, is prohibitively large and is expected to scale as the Coulomb interaction energy $\Delta E_{IT}$ between the ion and the electron~\cite{Tracy2014}. In the other limit, where the reaction complex is at its reduced state $\left(x,\xi\right)=\left(x_R,\xi_R\right)$, again, there is a large energy barrier to separate the electron from the transferred ion $\left(x,\xi\right)=\left(x_O,\xi_R\right)$ due to their electrostatic attraction. In other words, $\Delta E_{IT}$ corresponds to the energy required to eliminate an ion-electron pair (break local electroneutrality). This picture leads to the additional conditions for $f_O$ and $f_R$, which is
\begin{align}
    f_O\left(x_O,\xi_R\right) \simeq f_R\left(x_R,\xi_O\right) = \Delta E_{IT}
\label{eq:f_O_xo_f_R_xr}
\end{align}
Therefore, we argue that a sequential process would require supplying a considerable amount of energy in order to sacrifice the stabilizing attraction along that reaction pathway. 

Assuming the barrier is much larger than the thermal energy $k_BT$, the saddle point approximation for the first passage time \cite{risken1996fokker} is used to derive the rate at the point where $\left.\partial\mu_{O/R}/\partial\xi\right\rvert_{x_\ddag,
\xi_\ddag}=0$ (yellow point on the orange line) where the reaction barrier is minimized. At $(x_\ddag,\xi_\ddag)$, the electrostatic penalty for separating the ions and electrons from their final state is postulated to be approximately the same, eq.~\ref{eq:f_O_xo_f_R_xr}, leading to an additional constraint to the functional forms of $f_O$ and $f_R$
\begin{align}
    f_O\left(x_\ddag,\xi_\ddag\right) \simeq f_R\left(x_\ddag,\xi_\ddag\right)
\label{eq:f_O_xddag}
\end{align}
In cases where either the species do not interact through electrostatics, or the difference between $\mu_O$ and $\mu_R$ is much larger than the stabilizing electrostatic attraction $\Delta E_{IT}$, eq.~\ref{eq:f_O_xddag} might need to be modified. In such situation, `asymmetry' in the ionic coordinate can exist, and $f_O$ differs from $f_R$ at the transition state point $\left(x_\ddag,\xi_\ddag\right)$. The true nature of $f_O$ and $f_R$ can be revealed through detailed \textit{ab-initio} studies~\cite{Hartnig2003}.

The oxidized cluster has to fluctuate on a 2D energy surface until both the reactant and product states are equally likely to exist energetically, while the fluctuating trajectory is most likely to follow the path that passes over the minimum energy barrier for the reaction to happen. The point of the minimum transition state barrier $\left(x_\ddag,\xi_\ddag\right)$ is defined by 
\begin{equation}
    \mu_{O}^{ex}\left(x_\ddag,\xi_\ddag\right)=\mu_{R}^{ex}\left(x_\ddag,\xi_\ddag\right)=\mu_{\ddag}^{ex}\left(x_\ddag,\xi_\ddag\right)
\label{eq:CIET_TS}
\end{equation}
Additional terms that account for the widening of the RedOx species' density of states upon its interaction with the electrode~\cite{schmickler1986} can be added to eqs.~\ref{eq:mu_O_mrc} and~\ref{eq:mu_R_mrc}. 

The transition state chemical potential is split in two parts: 1) one that describes the polarization coordinate from traditional electron transfer kinetics $\mu_{\ddag,ET}^{ex}$, and 2) another which takes into account ionic effects of the TS landscape $f(x_\ddag,\xi_\ddag)$, e.g. surface site exclusion due to surface crowding phenomena~\cite{bazant2013,ferguson2012}. By assuming symmetric (equal) force constants for the polarization of the initial and final states $\kappa_O=\kappa_R=\kappa$, the functional form of the ET contribution is found after solving eqs.~\ref{eq:CIET_TS} for $x_\ddag$, as
~\cite{sutin1983,marcus1993}
\begin{equation}\label{eq:chem_ex_et}
\begin{split}
    \mu_{\ddagger,ET}^{ex} &= \mu_O^{ex}\left(x_O,\xi_O\right) + \frac{\lambda}{4}
    {\left( 1 + \frac{\mu_R^{ex}\left(x_R,\xi_R\right) - \mu_O^{ex}\left(x_O,\xi_O\right)}{\lambda} \right)}^2
    \\
    &= \mu_R^{ex}\left(x_R,\xi_R\right) + \frac{\lambda}{4}
    {\left( 1 - \frac{\mu_R^{ex}\left(x_R,\xi_R\right) - \mu_O^{ex}\left(x_O,\xi_O\right)}{\lambda} \right)}^2
\end{split}
\end{equation}
with reorganization energy $\lambda=\frac{\kappa}{2}\left(x_O-x_R\right)^2$ corresponding to the energy required to alter the environment of the oxidized/reduced state to that of the reduced/oxidized sate without allowing an electron transfer to occur. After substituting eq.~\ref{eq:f_O_xddag} and eq.~\ref{eq:chem_ex_et} in one of the equations in eq.~\ref{eq:mu_OR_mrc}, the final form of $\mu_{\ddag}^{ex}$ reads
\begin{align}
    \mu_{\ddag}^{ex} &= \mu_{\ddag,ET}^{ex} + f\left(x_\ddag,\xi_\ddag \right) = \mu_{\ddag,ET}^{ex} + \mu_{IT}^{ex}
    \label{eq:TST_muex}
\end{align}
where $\mu_{IT}^{ex}=\alpha_\xi\Delta E_{IT} + k_\textrm{B}T\ln\gamma_\ddag$. The first term in $\mu_{IT}^{ex}$ is the ion transfer barrier due to the electrostatic penalty on separating the ion from the electron, and the second term describes the ionic non-idealities on the transition state, such as excluded volume effects and activation strain energies~\cite{bazant2013}. The parameter $\alpha_\xi$ can be viewed as an ionic transfer coefficient, the value of which is determined by the exact functional form of $f_O$ and $f_R$, as we describe in Sec.~\ref{sec:BV_der}. In the case where the transferred ion is strongly coupled with the final position of the transferred electron, we expect the constants $\kappa_{O/R}$ to be a function of the ionic coordinate $\xi$. This can induce asymmetries between the RedOx parabolas in the polarization coordinates. Although, this consideration may be more realistic, it poses analytical difficulties on arriving to a simple analytical form for the transition state barrier $\mu_{\ddag}^{ex}$. 

\subsection{Quantum tunneling of electrons}\label{sec:ET}

The tunneling of electrons is modeled using the Landau-Zener theory~\cite{landau1932theorie,zener1932non}. The oxidized and reduced states are coupled to the surrounding reaction media which may fluctuate, leading to a crossing event of the wavefunctions at which tunneling may occur. The probability for such an event allowing for multiple re-crossings is given by~\cite{kuznetsov_book}
\begin{align}
    \textrm{P}\left(ET|O^{\ddag}/R^{\ddag}\right)&=k_T = \frac{1-\exp \left( -2 \pi \Gamma_{LZ} \right)}{1-\frac{1}{2}\exp \left( -2 \pi \Gamma_{LZ} \right)}; 
    \label{eq:Plz}
    \\
     \Gamma_{LZ} &= \left.\frac{H_{DA}^2}{h v_\ddag} \left( \frac{\partial |\mu_{O}^{ex}-\mu_{R}^{ex}|}{\partial x} \right)\right\rvert_{x_\ddag,\xi_\ddag}^{-1}
     \label{eq:gamma}
\end{align}
where, $k_T$ stands for the electron tunneling probability, $H_{DA}$ is the electronic coupling between the electron donor and acceptor, $v_\ddag$ is the thermally averaged reaction coordinate `velocity', $m$ is the effective mass of the reaction complex, and $h$ is Planck's constant~\cite{kuznetsov_book}. Because the coupling involves an electrode with electrons/holes of a manifold of energy levels, the electronic coupling is in general a function of electron energy levels $\varepsilon$. For non-adiabatic ET the weak coupling limit, $\Gamma_{LZ} \ll 1$, we obtain the more classical result~\cite{landau1932theorie,zener1932non,kuznetsov_book}
\begin{align}
    k_T = \frac{H_{DA}^2}{\nu_\ddag \hbar \sqrt{4 \pi \lambda k_{B} T }}
    \label{eq:Plz_nof}
\end{align}
where $\nu_\ddag$ is the frequency of the RedOx species along the harmonic reaction coordinate~\cite{kuznetsov_book} and $\hbar = h/2\pi$.

\subsection{ Electrostatic effects on ion transfer}\label{sec:solvation}
The main idea behind coupled ion-electron kinetics is the concerted transfer of both ions and electrons along the reaction coordinates. This process is mainly controlled by the interaction between ions and electrons that is described by the energy $\Delta E_{IT}$ found in eq.~\ref{eq:TST_muex}. Marcus in his original papers connected the reorganization energy with the Born solvation energy~\cite{marcus1965electron,marcus1993,kuznetsov_book}, giving a simple estimate of the electron solvation energy in a dielectric medium. Herein, we develop an analogous formula to estimate $\Delta E_{IT}$ based on electrostatics.
\begin{figure}[!ht]
    \centering
    \includegraphics[width=2in]{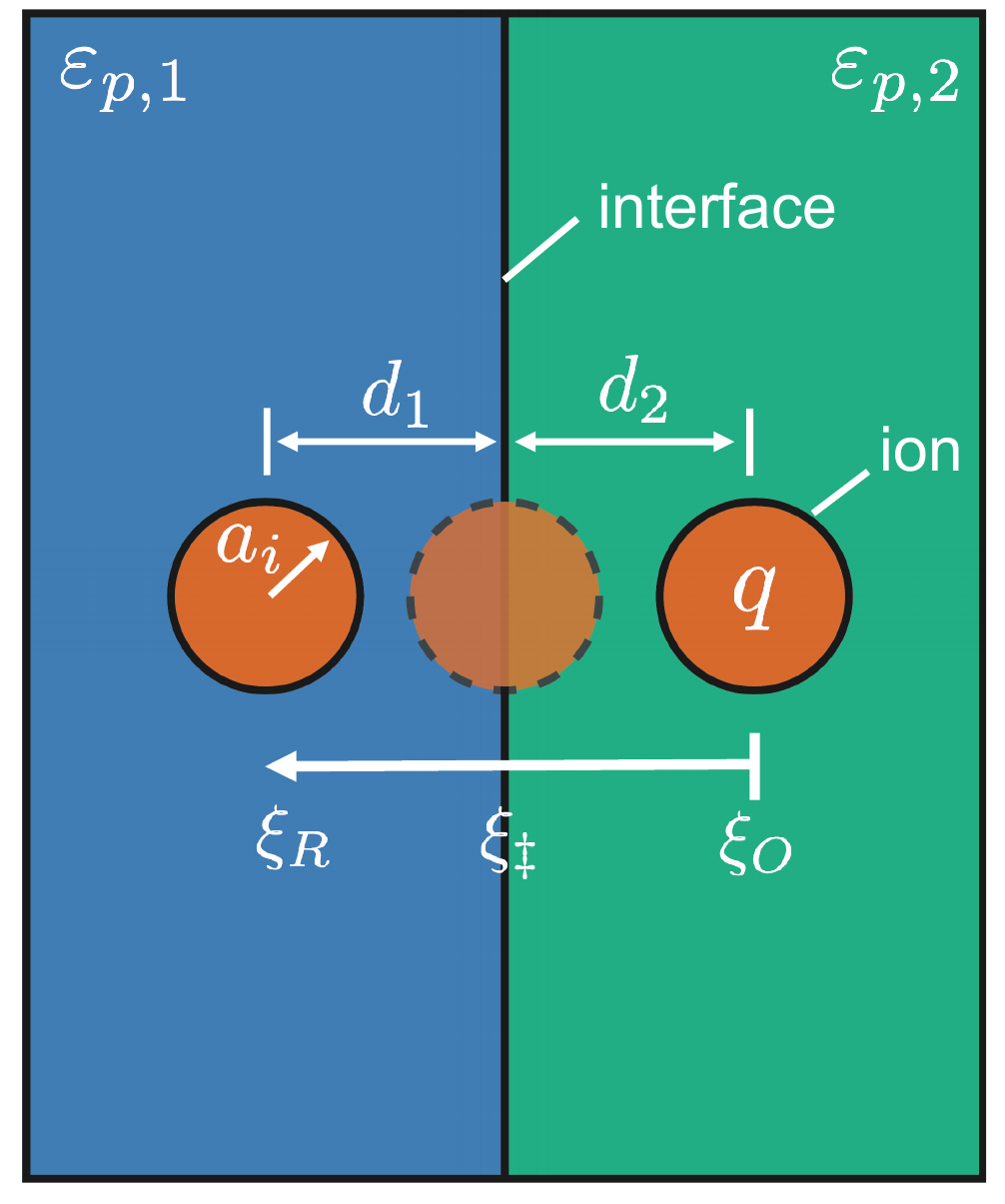}\hspace{8pt}
    \caption{Primitive electrostatic model for the solvation of an ion near an interface that connects two media with dielectric permittivities $\varepsilon_{p,1}$ and $\varepsilon_{p,2}$, respectively. The ion has charge $q$ and radius $a_i$. When the ion is solvated left/right of the interface, its distance from the interface is $d_{1/2}$. At the transition state $\xi_\ddag$, the ion is solvated exactly at the interface, where half of it is in the left dielectric medium and the other half on the right one. This model is similar to the one presented by Makov \& Nitzan~\cite{makov1994solvation}.}
    \label{fig:dielectric}
\end{figure}

We focus on the process $\left(x_R,\xi_R\right)\rightarrow\left(x_R,\xi_O\right)$. In this case, the electron and the ion are in the reduced state where we supply energy $\Delta E_{IT}$ to move the ion to the position it occupies when the reaction complex is in its oxidized state $\xi_O$. Therefore, we define $\Delta E_{IT}$ as
\begin{equation}
    \Delta E_{IT} \equiv \Delta G_{\xi_R\rightarrow\xi_O}
\label{eq:desolv_IT}
\end{equation}
where $\Delta G_{\xi_R\rightarrow\xi_O}$ includes the energy to separate the electron and ion pairs as well as desolvate and resolvate the ion between different media, like in the case of ion intercalation. For the solvation part of $\Delta G_{\xi_R\rightarrow\xi_O}$, we follow a similar analysis to that presented by Makov \& Nitzan~\cite{makov1994solvation}, who studied the effects of dielectric mismatch on the solvation of ions with finite size. 

According to fig.~\ref{fig:lndscpe}(a) and~\ref{fig:lndscpeBV}(a), the energy $\Delta G_{\xi_R\rightarrow\xi_O}$ is split in three contributions: 1) electrostatic interaction between the ion and the electron $E_C$, 2) desolvation of the ion from $\xi_R$ in vacuum, $\Delta G_{\xi_R\rightarrow v}$, 3) solvation of the ion from vacuum in $\xi_O$, $\Delta G_{v\rightarrow\xi_O}$. This is summarized as
\begin{equation}
    \Delta E_{IT} = E_C + \Delta G_{\xi_R\rightarrow  v} + \Delta G_{v\rightarrow \xi_O} 
\label{eq:solv_cont}
\end{equation}
where we need to supply $E_C$ in order to separate the ion-e$^-$ pair (break of electroneutrality). We expect this term to be the dominant one in the expression above, and can be approximated analytically using either the Coulomb or the screened Coulomb potential (Yukawa)~\cite{israelachvili2011intermolecular}
\begin{equation}
E_C =  \ 
 \left\{ \begin{array}{ll} 
 \  \ \frac{e^2}{4\pi\varepsilon_0\varepsilon_r r} & \ \  \mbox{Coulomb} \\
\ \ \frac{e^2}{4\pi\varepsilon_0\varepsilon_r r}e^{-r/\lambda_s} &\ \ \mbox{Screened Coulomb}
\end{array} \right.
\label{eq:dimless_criterion}
\end{equation}
where $\varepsilon_0$ and $\varepsilon_r$ are the permittivities of vacuum and the dielectric medium, $\lambda_s$ is the screening length, and $r$ the distance between the localized electron and the ion.

For the solvation process, both $\Delta G_{\xi_R\rightarrow v}$ and $\Delta G_{v\rightarrow \xi_O}$ depend on the ion radius $a_i$, the permittivities of the dielectric media $\varepsilon_{p,1}$, $\varepsilon_{p,2}$, and on the distances $d_{1}$ and $d_{2}$ of the ion from the interface – 1 stands for the electrolyte phase and 2 for the solid phase, fig.~\ref{fig:dielectric}. Following~\cite{makov1994solvation}, we use eqs. 17 in their work to derive the following form of $\Delta E_{IT}$
\begin{equation}
\label{eq:DeltaEIT_int}
\begin{split}
\Delta E_{IT} = & 
    E_C + \frac{e^2}{2}\left[\frac{1}{a_i}\left(\frac{1}{\varepsilon_{p,1}}-\frac{1}{\varepsilon_{p,2}}\right)+\right. \\
    & \left.\frac{1}{2}\left(\frac{\varepsilon_{p,2}-\varepsilon_{p,1}}{\varepsilon_{p,1}+\varepsilon_{p,2}}\right)
    \left(\frac{1}{d_2\varepsilon_{p,2}}-\frac{1}{d_1\varepsilon_{p,1}}\right)\right]
\end{split}
\end{equation}
When $\xi_O$ and $\xi_R$ correspond to the the same physical position, then the second term in Eq.~\ref{eq:DeltaEIT_int} is zero.

One can directly consider the relation between the energy contributions of the transition state due to the solvation effects and $\Delta E_{IT}$. {In this case, the two dielectric media are not in equilibrium as the transferring ion is at $\xi=\xi_\ddag$, so $\varepsilon_{p,1}$ and $\varepsilon_{p,2}$ correspond to non-equilibrium permittivities~\cite{kuznetsov_book}. As a first approximation the solvation energy of the transferring ion at the transition state, we assume that its the environment is at equilibrium, and thus $\varepsilon_{p,i}$ are the static permittivities of the two media.} As discussed earlier, the saddle point approximation predicts that for $\xi=\xi_\ddag$ the ion is midway to its final state. One can assume this position to be right on top of a fictitious interface between the two dielectric media where the ion is solvated. Again, using eq. 18 from Ref.~\cite{makov1994solvation}, we can describe the electrostatic part of the transition state as a solvation process from the vacuum state to the interface between the two dielectrics. Thus, we find a simple correspondence between $\Delta E_{IT}$ and the electrostatic part due to solvation of the transition state energy as follows
\begin{equation}
\label{eq:DeltaEIT_int_interface}
\begin{split}
\Delta E_{IT} = 
    E_C - \frac{e^2}{2\alpha_\xi a_i}\left(\frac{\varepsilon_{p,1}+\varepsilon_{p,2}-2}{\varepsilon_{p,1}+\varepsilon_{p,2}}\right)
\end{split}
\end{equation}
Based on this form, it is apparent that by tuning the dielectric mismatch between the two media, one can promote or suppress the reaction dynamics. Eq.~\ref{eq:DeltaEIT_int_interface} can be seen as an extrapolation from the transition state to the final state of the system -- the oxidized or reduced one. 

The relations presented are simple approximations to the real system. The actual $\Delta E_{IT}$ can be calculated by performing either molecular dynamics~\cite{Hartnig2003} or \textit{ab-initio} simulations, where the structure of the medium as well as the dynamics of the reactant species are taken into account in a systematic way. For example, Maxisch et al.~\cite{Maxisch2006_polaron} estimated $\Delta E_{IT}\simeq E_C = 0.37eV$ for the Li$^+$-polaron interaction in a LFP crystal.

\subsection{Rate of coupled ion-electron transfer }\label{sec:current}
Combining eq.~\ref{eq:chem_ex_et} with eq.~\ref{eq:TST_muex} and using the definition of the formal overpotential and overpotential of eqs.~\ref{eq:formal_overpotential}-\ref{eq:overpotential}, as well as the tunnelling probability eq.~\ref{eq:Plz_nof} in the elementary reaction rate expressions, eq.~\ref{eq:R_rxn_fwd_bck_full}, we arrive at
\begin{widetext}
\begin{subequations}
\label{eq:R_total_final}
\begin{equation}\label{eq:R_fwd}
R_{red,\varepsilon} = 
    \frac{\tilde{k}_0 e^{-\alpha_\xi\Delta E_{IT}/k_BT}}{\gamma_{\ddag}} {c}_O n_e
    \exp\left( -\frac{{\left( \lambda + e\eta_f - \left(\varepsilon - E_f\right) \right)}^2}{4\lambda k_{B}T} \right)
\end{equation}
\begin{equation}\label{eq:R_bck}
R_{ox,\varepsilon} = 
    \frac{\tilde{k}_0 e^{-\alpha_\xi\Delta E_{IT}/k_BT}}{\gamma_{\ddag}} {c}_R \left(1-n_e\right)
    \exp\left( -\frac{{\left( \lambda - e\eta_f + \left(\varepsilon - E_f\right) \right)}^2}{4\lambda k_{B}T} \right)
\end{equation}
\end{subequations}
\end{widetext}
where $E_f$ is the Fermi level of the electron donor and $\tilde{k}_0$ is the lumped reaction rate prefactor
\begin{equation}
    \tilde{k}_0 = k_0k_T e^{-w_{O/R}/k_BT}
    \label{eq:w_or}
\end{equation}
The electron concentration $n_e=1/\left(1+e^{\varepsilon-E_f}\right)$ is described by the Fermi-Dirac distribution~\cite{SchmicklerText,kuznetsov_book}. {In both rate expressions, we defined the formal overpotential $\eta_f$, which is a measure of the departure of the electrode potential from the formal one as a result of RedOx concentration effects~\cite{kuznetsov_book}, as
\begin{equation}\label{eq:formal_overpotential}
    e\eta_f = e\eta + k_BT\ln\left(\frac{c_O}{c_R}\right)
\end{equation}
where $\eta$ is the overpotential {defined as~\cite{bazant2013} 
\begin{equation}
    e\eta = \mu_R-\mu_O -\mu_e = eV^\Theta + k_B T\ln\left(\frac{\gamma_R c_R}{\gamma_Oc_O}\right) + e\left(z_R\phi_R-z_O\phi_O\right)
    \label{eq:overpotential}
\end{equation}} 
and the formal potential of the reaction is defined as $eV^\Theta=\mu_R^\Theta - \mu_O^\Theta - E_f$. Last, to arrive at the final form of eqs.~\ref{eq:R_total_final}(a) \& (b), we considered the excess chemical potential of the electrons to be $\mu^{ex}_e=\varepsilon=\epsilon-e\phi_e$, and also $\mu_e=E_f=\varepsilon + k_BT\ln\left(n_e/(1-n_e)\right)$. 

We can connect the form of eqs.~\ref{eq:R_total_final} with those found in classical electron transfer papers~\cite{chidsey1991} and electrochemistry books~\cite{bard2001}. In connection to the notation used in Chidsey's paper on electron transfer reactions at metal-electrolyte interfaces~\cite{chidsey1991}, $x\equiv\left(\varepsilon-E_f\right)/k_BT$ and $E^{0'}-E\equiv e\eta_f$, where $E^{0'}$ and $E$ are the formal and electrode potential, respectively. Related to electrochemistry books like Bard \& Faulkner~\cite{bard2001}, the formulation of eqs.~\ref{eq:R_total_final} can be directly translated to the expressions of eqs. (3.6.34) \& (3.6.35) (p. 129-130) under standard conditions where $e\eta_f= e\eta= \mu_R^\Theta - \mu_O^\Theta - E_f$. Thus, performing the change of variables $x\equiv E-E_F$ in eqs. (3.6.34) \& (3.6.35) of~\cite{bard2001}, we arrive at the same form for the forward and backward reaction rates, where $E$ and $E_F$ are the electron and Fermi energies, respectively.} 


In eqs.~\ref{eq:R_total_final}, the term $e^{-\alpha_\xi\Delta E_{IT}/k_BT}/\gamma_{\ddag}$ is the main contribution of this work, where it describes the ion transfer effects on the transition state coupled with electron transfer. The other contributions such as the electron transfer term, the work required to bring the species to their RedOx states, and the tunneling factor are common in the field of electron transfer~\cite{kuznetsov_book,marcus1993,fedorov2014}. Finally, we are interested in validating the ionic dependencies on the transition state encoded in $\gamma_\ddag$. We do so by applying the developed theory in ion intercalation. Thus, we go one step further and absorb the $e^{-\alpha_\xi\Delta E_{IT}/k_BT}$ into the reaction rate prefactor as 
$$k_0^*=\tilde{k}_0e^{-\alpha_\xi\Delta E_{IT}/k_BT}.$$ 
For the remainder of the paper, all energetic quantities are normalized to $k_BT$.


\section{Two Limits Leading to the Butler-Volmer Equation }\label{sec:BV_der}

The ionic coordinate is tightly connected with the Coulomb energy which is a result of the attraction between ions and electrons, and thus $f_O$ and $f_R$ scale with $\Delta E_{IT}$. From classical Marcus theory~\cite{marcus1993}, it is known that the electron transfer energy is proportional to the reorganization energy $\lambda$, which in our model is expressed in terms of the curvature of the parabolas, $\kappa$. Given that we have two characteristic energy scales, dimensionless analysis shows that their ratio $\kappa/\Delta E_{IT}$ serves as a characteristic measure, and helps us understand the limits of the developed model.

\subsection{Electron-transfer limitation}

The first case is that of $\kappa/\Delta E_{IT}\gg1$. In this scenario, ion transfer is decoupled from electron transfer ($f_{O/R}/\kappa\rightarrow0$), and thus the reaction is limited by the environment reorganization and electron tunneling only. The reaction is solely described in the solvent polarization coordinate and the classical approach of the two intersecting parabolas is followed. As a result, we recover the original Marcus/MHC model, where only the electron transfer needs to be resolved, while ion transfer dependencies are lumped in the constants of the model. For overpotential values smaller than $\lambda/e$, one arrives at the BV model with charge transfer coefficient $\alpha=1/2$. This result is well-known~\cite{bard2001,bazant2013} and serves as the classical approach for providing a physical picture to the phenomenological Butler-Volmer kinetics.

Another interesting limit is when the ratio $e\eta/\lambda\gg1$. In this case, one finds that $\mu_{\ddag,ET}^{ex}\propto \eta^2$ and therefore the activation energy barrier scales with $E_{act} \sim -\eta^2/4\lambda k_BT$. For localized electrons, the resulting reaction rate is predicted to decrease with increasing imposed driving force. This behavior leads to the well-known Marcus inverted region~\cite{marcus1993,kuznetsov_book}.

\subsection{Ion-transfer limitation}

\begin{figure*}[!ht]
    \centering
    \includegraphics[width=1\textwidth]{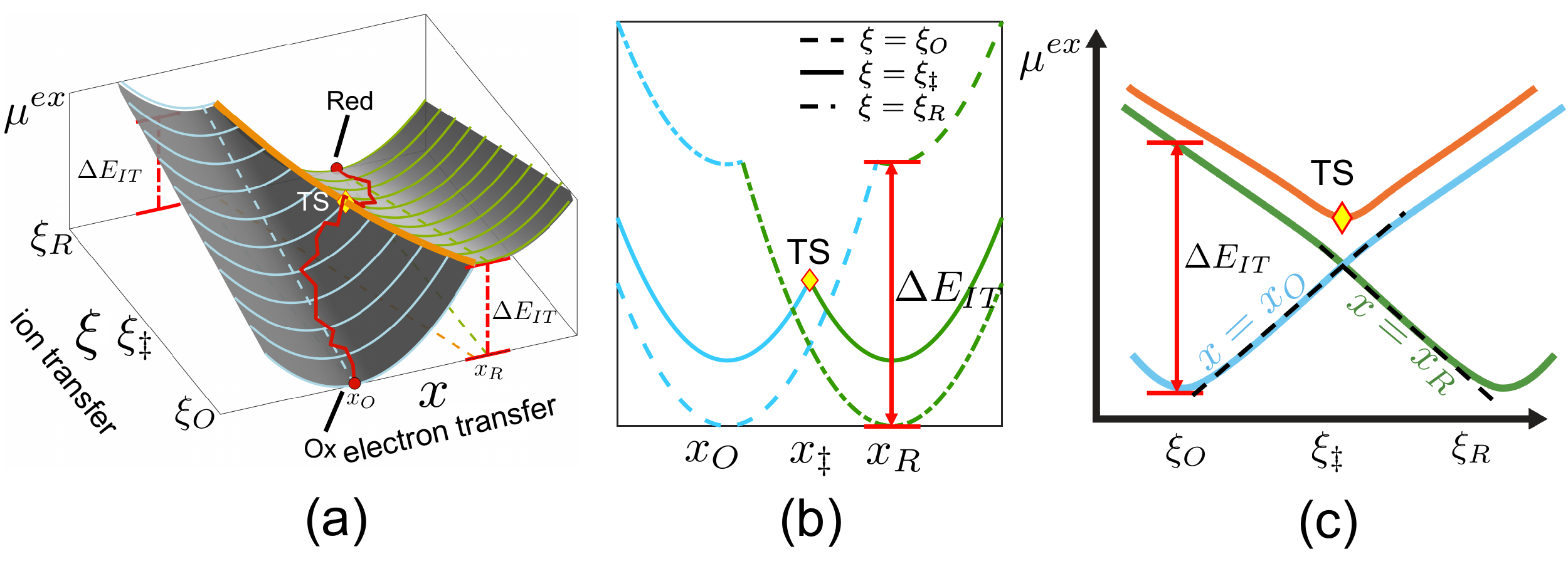}\hspace{8pt}
    \caption{(a) Excess chemical potential landscape for both reactants and products vs. the reaction coordinates $\left(x,\xi\right)$, when the dependence in $\xi$ is linear. The description of the lines and the fluctuating path in the energy landscape is given in fig.~\ref{fig:lndscpe}. (b) Excess chemical potential landscape for $\xi=\xi_O$ (dashed line), $\xi=\xi_\ddag$ (solid line), $\xi=\xi_R$ (dashed-dot line). (c) Schematic illustration of the projection of the energy profiles of the oxidized ($x=x_O$) and reduced ($x=x_R$) species on the energy-$\xi$ plane is shown.}
    \label{fig:lndscpeBV}
\end{figure*}

In the limiting case of $\kappa/\Delta E_{IT}\ll1$, which can be due to steep changes in the electrostatic potential nearby the electrode, e.g. diffuse double layers~\cite{bazant2005current,chu2005electrochemical,Limaye2020}, the analysis shows an interesting connection between coupled ion-electron transfer and Butler-Volmer kinetics~\cite{butler1924part3,erdey-gruz1931zur,bard2001}. Under these conditions, we consider the dependence of $f_O$ and $f_R$ to be linear in the ionic coordinate $\xi$ leading to the following expressions for the RedOx species chemical potentials
\begin{subequations}
\begin{align}
    \mu_{O}^{ex}(x,\xi) \simeq \mu_{O}^{ex}\left(x_O,\xi_O\right) + \frac{\kappa}{2}\left(x-x_O\right)^2  + \Delta E_{IT} \xi
\label{eq:mu_O_mrc_lin} \\
    \mu_{R}^{ex}(x,\xi) \simeq \mu_{R}^{ex}\left(x_R,\xi_R\right) + \frac{\kappa}{2}\left(x-x_R\right)^2  + \Delta E_{IT} \left(1-\xi\right)
\label{eq:mu_R_mrc_lin}
\end{align}
\label{eq:mu_R_mrc_lin_both}
\end{subequations}
as also shown in Fig.~\ref{fig:lndscpeBV}. By following the classical procedure on finding the intersection of $\mu_O$ and $\mu_R$ in the $x$ coordinate, eq.~\ref{eq:CIET_TS}, the transition state line $x_\ddag$ as a function of $\xi$ is
\begin{widetext}
\begin{equation}
    x_\ddag\left(\xi\right) = \frac{1}{2} \left(x_R+x_O + \frac{2 \left(\Delta E_{IT}(1-2\xi) + \mu_{R}^{ex}\left(x_R,\xi_R\right) - \mu_{O}^{ex}\left(x_O,\xi_O\right) \right)}{\kappa (x_R-x_O)}\right)
    \label{eq:x_ddag_linear}
\end{equation}
\end{widetext}
Substituting back to $\mu_O$ or $\mu_R$, eq.~\ref{eq:mu_OR_mrc}, and minimizing over $\xi$ we find the following explicit expression for $\xi_\ddag$
\begin{equation}
    \left.\frac{\partial \mu_O^{ex}}{\partial \xi}\right\rvert_{x_\ddag,\xi_\ddag} = 0 \rightarrow \xi_\ddag = \frac{\Delta E_{IT}+\mu_{R}^{ex}\left(x_R,\xi_R\right) - \mu_{O}^{ex}\left(x_O,\xi_O\right)}{2\Delta E_{IT}}
    \label{eq:xi_linear_sol}
\end{equation}
Finally, the transition state chemical potential, which corresponds to the minimum energy barrier for the reaction to proceed, is found by substituting eq.~\ref{eq:xi_linear_sol} and eq.~\ref{eq:x_ddag_linear} in eq.~\ref{eq:mu_R_mrc_lin_both}
\begin{equation}
\begin{split}
    \mu_\ddag^{ex} = \frac{1}{2}\left(\Delta E_{IT}+\mu_{R}^{ex}\left(x_R,\xi_R\right) + \mu_{O}^{ex}\left(x_O,\xi_O\right) + \frac{\lambda}{2}\right)
\end{split}
\label{eq:CIET_linear_TST}
\end{equation}
where for $\kappa/\Delta E_{IT}\ll1$, Fig.~\ref{fig:lndscpeBV}(b), we can drop the $\lambda/2$ term). In this limit, we find $\mu^{ex}_\ddag$ to be equivalent to the transition state of Butler-Volmer kinetics~\cite{bazant2013} for $\alpha=1/2$, where also we can see that $\alpha_\xi=\alpha$. This is understandable in retrospect since the two-dimensional energy landscape can be cast in a form which imitates the phenomenological charge-transfer coordinate used in deriving Butler-Volmer kinetics, Fig.~\ref{fig:lndscpeBV}(c). Our analysis reveals a different mechanism that recovers Butler-Volmer kinetics via coupled ion-electron transfer, and is in agreement with Fokker-Planck approaches commonly used in the field of quantum chemistry~\cite{nitzan2006chemical,Limaye2020}.

\section{Application to Electrodes} 

\subsection{Faradaic current}

In order to obtain the total Faradaic reaction rate at an electrode, we assume a continuum of electron energy levels with density of states $\rho\left(\varepsilon\right)$, which corresponds to a family of parabolas in the electron transfer coordinate $x$~\cite{zeng2014simple,bai2014,kuznetsov_book,SchmicklerText,henstridge2011marcus,laborda2013asymmetric}. By integrating eq.~\ref{eq:R_rxn_P} over all available energy levels we arrive at
\begin{align}
\label{eq:R_tot}
R = R_{red} - R_{ox} =\int\limits_{-\infty}^{\infty} \left(R_{red,\varepsilon}-R_{ox,\varepsilon}\right) \rho\, d \varepsilon
\end{align}
where in the weak coupling limit $\Gamma_{LZ} \ll 1$, eq.~\ref{eq:gamma}, $R_{red,\varepsilon}$ and $R_{ox,\varepsilon}$ depend on the electron energy level $\varepsilon$ through $n_e$. {The forward $R_{red}$ and backward $R_{ox}$ reaction rates satisfy the De Donder relation $R_{red}/R_{ox}=e^{-e\eta/k_BT}$, in compliance with microscopic reversibility~\cite{kondepudi2014modern,bazant2013}. Details on the derivation of the De Donder relation for coupled ion-electron transfer kinetics are given in the Appendix.}

It is convenient to recast the net reaction rate in current density form, $i=eR$, as a function of the overpotential, as defined in eq.~\ref{eq:overpotential}, and a prefactor defining the exchange current density~\cite{bazant2013},
\begin{align}
    i
    = \int_{-\infty}^{\infty} i_0 
     \left[ e^{-\alpha \eta }
    - e^{\left(1-\alpha\right) \eta } \right] \rho \, d \varepsilon
\label{eq:itot_bv}
\end{align}
{After some lengthy algebra for factorizing the current density in the form of Butler-Volmer, we arrive at the following form for the exchange current density
\begin{widetext}
\begin{align}
\begin{split}
    i_0(\varepsilon,\eta,c_i) = &\frac{ek^*_0}{\gamma_\ddag}
    e^{-\frac{\left(\varepsilon-E_f-\lambda\right)^2}{4\lambda}}
    {c_O}^{\frac{(3-2\alpha+(\varepsilon-E_f)/\lambda)}{4}}{c_R }^{\frac{(1+2\alpha-(\varepsilon-E_f)/\lambda)}{4}}n_e e^{-\frac{\eta^2}{4 \lambda}} 
    = \frac{ek^*_0}{\gamma_\ddag} c_On_e e^{-\alpha^2\lambda}e^{-\frac{\eta^2}{4 \lambda}}
\label{eq:i0}
\end{split}
\end{align} 
\end{widetext}
and we define the charge transfer coefficient as
\begin{align}
    \alpha\left(\varepsilon\right) = \frac{1}{2}\left( 1 + \frac{1}{\lambda}\ln{\left( \frac{c_O}{c_R} \right)}
    + \frac{E_f-\varepsilon}{\lambda} \right).
\end{align}

In eq.~\ref{eq:i0}, we observe that when ion transfer limitations are negligible, then $\gamma_\ddag\simeq1$ and $k_0^*\simeq k_0k_Te^{-w}$ and the exchange current density is equal to
\begin{align}
\begin{split}
    i_0(\varepsilon,\eta,c_i) =  
    & \frac{ek_0\widetilde{H}_{DA}^2}{\sqrt{4 \pi \lambda }} e^{-w} c_O n_e e^{-\alpha^2\lambda}e^{-\frac{\eta^2}{4 \lambda}}
\label{eq:i0_ET_limited}
\end{split}
\end{align} 
where $\widetilde{H}_{DA} = H_{DA}/\sqrt{\nu_\ddag \hbar k_BT}$ is the dimensionless energy barrier between the two non-adiabatic electron states of the RedOx species~\cite{zwanzig2001nonequilibrium} and we considered $w_O\simeq w_R\equiv w$. Eq.~\ref{eq:i0_ET_limited} in combination with eq.~\ref{eq:itot_bv} are the BV-like form of the Marcus-Hush-Chidsey model, which simplifies into the Marcus model~\cite{bazant2013} when the density of states corresponds to a single electron level. 

In BV models, the value of $i_0$ is estimated at $\eta\rightarrow0$~\cite{bard2001}. Here, we see that $i_0$ is within the integral related to the available energy levels of the electron donor. To be consistent with the electrochemistry literature, the classic exchange current density $\widetilde{i}_0$ is defined as 
\begin{equation}\label{eq:exch_cur_dens_formal}
    \widetilde{i}_0 = \lim_{\eta\rightarrow0} \int_{-\infty}^{\infty} e R_{red} \rho\,d\varepsilon = \lim_{\eta\rightarrow0} \int_{-\infty}^{\infty} e R_{ox} \rho\,d\varepsilon = \int_{-\infty}^{\infty} i_0 \rho\,d\varepsilon 
\end{equation}
The integral does not have an analytical form, except in specific cases, where for instance $n_e=\Theta(E_f-\varepsilon)$, with $\Theta$ to be the Heaviside step function or when uniformly valid approximations are used~\cite{zeng2014simple}. In both of these cases and under dilute conditions, the exchange current density can be written as 
\begin{equation}\label{eq:exch_current_density}
    \widetilde{i}_0 \simeq \frac{ek_0\widetilde{H}_{DA}^2}{2\gamma_\ddag} e^{-w} e^{-\alpha_\xi \Delta E_{IT}} c_Of\left(\lambda\right)
\end{equation}
where $f\left(\lambda\right)=\mathrm{erfc}\left(\frac{\sqrt{\lambda}}{2}-\frac{A}{2}\sqrt{\frac{1}{\lambda}+\frac{1}{\sqrt{\lambda}}}\right)$, with $A=0$ for $n_e=\Theta\left(E_f-\varepsilon\right)$ and $A=1$ for the uniformly valid approximation of~\cite{zeng2014simple}. When electron transfer near a metallic electrode is the limiting step (MHC model~\cite{chidsey1991}), the exchange current density becomes $\widetilde{i}_0 \simeq ek_0\widetilde{H}_{DA}^2 e^{-w} c_O f\left(\lambda\right)$. In the case where the reorganization energy $\lambda$ does not depend on species concentration, the form of $\widetilde{i}_0$ depends on $\lambda$ as shown in Fig.~\ref{fig:overpotential_dep}(a). For simplicity, we assume dilute solution and for completeness we include the approximated forms of eq.~\ref{eq:exch_current_density}.}

The model parameters can be obtained either through experiments or first-principle calculations. In particular, the chemical potential of the species can be found either using equilibrium statistical mechanical methods~\cite{balluffi2005kinetics}, or using experimental (non-)equilibrium measurements~\cite{sekimoto2010} such as construction of Tafel plots~\cite{bai2014}. The common practice for estimating the electron donor density of states $\rho(\varepsilon)$ and the reorganization energy $\lambda$ is by using density functional theory, via the calculation of the band and phonon structures of the materials used in the reaction~\cite{emin2013polarons}.

\subsubsection{Localized electrons}

In the case of localized electrons for an insulating electrode or isolated molecule, the density of states can be approximated by a Dirac delta function around the localized energy level $\varepsilon_0$ as $\rho=\delta(\varepsilon-\varepsilon_0)$. {Here, the Fermi level $E_f$ corresponds to the single electron level $\varepsilon_0$.} Additionally, the factors which account for the probability of finding occupied $n_e$ and unoccupied $1-n_e$ energy levels in eqs.~\ref{eq:R_fwd}-\ref{eq:R_bck} need to be set equal to $1$. In that case, the total reaction rate expression reads
\begin{align}
\label{eq:R_tot_dos_delta}
R =  \frac{k^*_0}{\gamma_{\ddag}}\left( c_Oe^{ -\frac{{\left( \lambda + e\eta_f \right)}^2}{4\lambda}}-
c_Re^{-\frac{{\left( \lambda - e\eta_f \right)}^2}{4\lambda} }\right)
\end{align}
where the second part of the expression corresponds to the electron transfer event, as has been initially derived by Marcus and others~\cite{marcus1993,kuznetsov_book}. The first part corresponds to the ionic part of the transition state according to the developed framework of coupled ion-electron transfer. A similar form of eq.~\ref{eq:R_tot_dos_delta} has been given in~\cite{bazant2013}, where the transition state activity coefficient $\gamma_\ddag$ has been included through a `modified' reorganization energy that depends on the RedOx species concentrations, e.g. $\lambda\left(c_O,c_R\right)$. 

\begin{figure*}[!ht]
    \centering
    \includegraphics[width=1\textwidth]{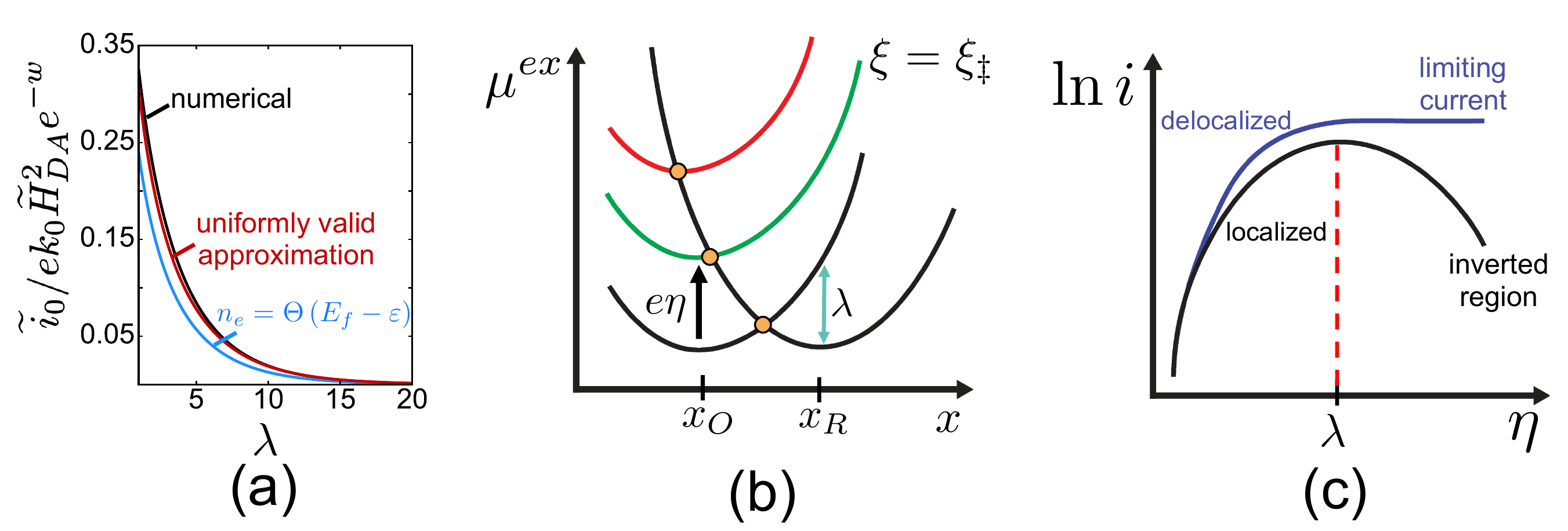}\hspace{8pt}
    \caption{(a) Normalized exchange current density in terms of the reorganization energy $\lambda$ in the case of dilute solution. For completeness we include the results of the numerical integration of eq.~\ref{eq:exch_cur_dens_formal}, the uniformly valid approximation of~\cite{zeng2014simple}, and the when $n_e=\Theta\left(E_f-\varepsilon\right)$. (b) Energy landscape in terms of the reorganization coordinate $x$ for constant $\xi=\xi_\ddag$ for different values of the overpotential. (c) Tafel plot for the case of localized (black) and delocalized (blue) donor electrons. In the case where the electrons originate from an insulating phase or an isolated molecule, CIET predicts the inverted region observed in the classical model by Marcus~\cite{marcus1993}, where the current decreases with increasing driving force for $\eta >\lambda$. Here, $\eta$ is scaled with $k_BT/e$ and $\lambda$ with $k_BT$. When the electron donor has delocalized electrons, a family of parabolas exists that leads to a reaction-limited current for $\eta >\lambda$. }
    \label{fig:overpotential_dep}
\end{figure*}

In Figs.~\ref{fig:overpotential_dep}, we present the reaction landscape in terms of $x$ at $\xi=\xi_\ddag$ and the current vs. overpotential dependence. In both figures, $\eta$ and $\lambda$ are scaled with $k_BT/e$ and $k_BT$, respectively. At the transition state point $\left(x_\ddag,\xi_\ddag\right)$, CIET predicts the classical picture of the energy landscape that Marcus reported in his seminal works~\cite{marcus1993,kuznetsov_book,SchmicklerText,bard2001}, where for $\eta=\lambda$ the electron transfer reaction becomes barrier-less~\cite{fraggedakis_stability2019}, Fig.~\ref{fig:overpotential_dep}(b). For $\eta>\lambda$, however, the electron transfer barrier starts increasing again leading to the Marcus inverted region~\cite{bazant2013,fraggedakis_stability2019}, Fig.~\ref{fig:overpotential_dep}(c). Finally, across the line where the parabolas intersect, there are values of $\xi$ for which $x_\ddag(\xi)$ lies in the inverted region and others that do not. For $\eta\gg\Delta E_{IT}$ though, the transition state point in the ionic coordinate becomes $\xi_\ddag=\xi_O$, resulting in negligible ionic barrier and the reaction becomes electron-transfer limited.

\subsubsection{Delocalized electrons}

When the electrons that participate in the electrochemical reaction come from a metal, the density of states $\rho$ is approximated as uniform nearby the Fermi level $E_f$. In that case, the total reaction rate is calculated by eq.~\ref{eq:R_tot} by setting $\rho$ to be constant. Generally, the resulting integral does not admit an analytical solution, except in certain limits where the Fermi-Dirac distribution can be approximated with the Boltzmann distribution~\cite{SchmicklerText}. Despite these difficulties, one can either use special quadrature rules to evaluate the integral with very few function evaluations, e.g. Gauss–Laguerre quadrature, or other analytical approximations with acceptable numerical accuracy. Here, we briefly review an analytical approximation to the integral over all available energy levels for the case of constant $\rho$~\cite{zeng2014simple}. 
After applying the approximation of eq. 17 of Ref.~\cite{zeng2014simple}, the total reaction rate becomes
\begin{equation}
\label{eq:approximation}
R = \frac{k_0^*\sqrt{\pi\lambda}}{\gamma_\ddag}\left(\frac{c_O}{1+e^{\eta_f}}
-
\frac{c_R}{1+e^{-\eta_f}}\right){\mathrm{erfc}\left( \frac{{\lambda -\sqrt{\hat{\alpha}+\eta_f^2}}}{2\sqrt{\lambda}} \right )} 
\end{equation}
where $\hat{\alpha}=1+\sqrt{\lambda}$. As discussed in~\cite{zeng2014simple}, Eq.~\ref{eq:approximation} can be evaluated as quickly as BV and does not require numerical integration, as implied from eq.~\ref{eq:R_tot}. This fact makes it convenient for its use in analytical models. Additionally, it was shown in fig.~4 of~\cite{zeng2014simple} that it is extremely accurate in various limits. More specifically, in the physically relevant case where $\lambda > k_BT$, the approximation error is always bounded below 5\% even for small values of $\eta_f$. At large $\eta_f$ and/or large $\lambda$ the formula is able to replicate with extreme accuracy the results obtained from numerical quadrature, since it has exponentially small error in both $\eta_f$ and $\lambda$. Therefore, we believe that in the case of a metallic electron donor, the integral appearing in eq.~\ref{eq:R_tot} can be approximated by the formula given in eq.~\ref{eq:approximation} with high accuracy and numerical efficiency.

In the case of delocalized electrons, the energy landscape of Fig.~\ref{fig:overpotential_dep}(b) would correspond to a family of parabolas for the oxidized state, instead of a single one. Thus for metallic electron donors, CIET predicts the current density to saturate for $\eta>\lambda$, Fig.~\ref{fig:overpotential_dep}(c), while its limiting value is affected by the ionic energy barrier, eq.~\ref{eq:TST_muex}.

\section{Application to Ion Intercalation}\label{sec:results}

\subsection{Motivation}

Intercalation is a reversible reaction of ion insertion and extraction, whereby the stoichiometry of the host material changes with increasing/decreasing ion concentration. The insertion of species may be driven chemically, e.g. hydrogen insertion in Pd~\cite{Hayee2018}, or electrochemically, e.g. Li ion intercalation in oxides~\cite{Nitta2015}. Ion intercalation has been traditionally modeled by charge transfer kinetics using the BV equation in the context of batteries~\cite{newman2004,doyle1993}. While charge transfer kinetics does not explicitly specify the nature of charge, which can be either that of the transferred electron or ion, it is often assumed that ion intercalation is limited by IT due to the experimentally observed concentration-dependent current densities~\cite{latz2013thermodynamic,nikitina2020metal,dreyer2016new}. Some have even questioned whether ion intercalation is a Faradaic reaction, arguing that ET from the electrode does not occur, beyond the simple electrostatic response of a capacitor~\cite{biesheuvel2018difference}.

\begin{figure}[!ht]
\centering
\includegraphics[width=3.2in]{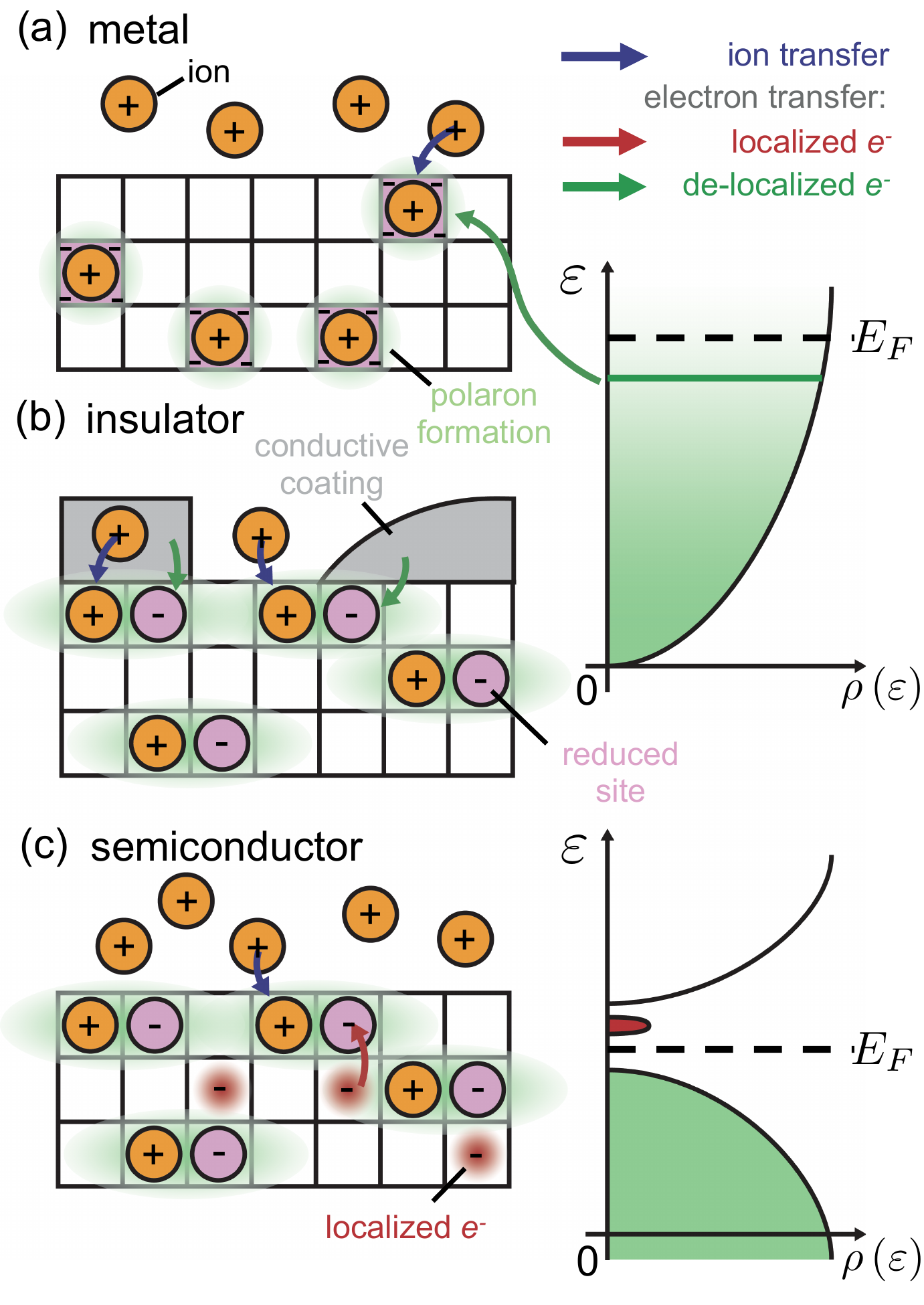}\label{fig:examples_e_band}\hspace{8pt}
\caption{Schematic illustration of the intercalation process described by coupled ion-electron transfer kinetics. The ion originates from a reservoir outside the particle, while the electron is provided by (a) \& (b) a metallic or (c) a semiconducting source. In the present figure, the electron $(e^-)$ donor corresponds to a metal or semi-conductor, with Fermi energy $E_F$. In (a) metals, electrons are delocalized within the solid, while in (c) semi-conductors they are localized on homogeneously distributed dopants. In the case of (b) an insulating material like LiFePO$_4$, the electrons are provided from a thin conductive coating.}
\label{fig:intercalation_in_out}
\end{figure}

Contradicting this paradigm, it was recently proposed that ion intercalation in the Li-ion battery cathode material, Li$_x$FePO$_4$ (LFP), is instead limited by electron transfer. In particular, it was argued that the Li$^+$ ion transfer step is fast and follows the slow transfer of electrons between the metallic carbon coating and the neighboring RedOx-active Fe$^{3+}$/Fe$^{2+}$ sites in the insulating crystal host~\cite{bai2014}. These developments suggest that both ET and IT are important in ion intercalation, and the coupling between them depends on the electronic nature of the host compound.  The materials in which ions intercalate can be metallic, semi-conducting or insulating. For example, LFP is a poor electronic conductor, while Li$_x$C$_6$ (graphite) acts as a metal. Both are common materials found in commercial Li-ion batteries. Figs.~\ref{fig:intercalation_in_out} illustrate the case of coupled ion-electron transfer applied to ion intercalation, where the electron donor might be conducting or semi-conducting. Initially, ions exist in the electrolyte reservoir, and electrons reside either in the environment of the system (e.g. in the LFP case, where it comes from the carbon coating, fig.~\ref{fig:intercalation_in_out}(b)), or in the solid matrix of the intercalation material (e.g. as in Li$_x$CoO$_2$ (LCO) and graphite, because of their metallic state, fig.~\ref{fig:intercalation_in_out}(a) \& (c)), or in a combination of both. 

\subsection{ CIET applied to ion intercalation }\label{sec:excluded_volume}

The proposed mechanism of coupled ion-electron transfer takes into account the effects of the ion environment on the transition state. In ion intercalation materials, a non-negligible phenomenon that affects both the thermodynamics and the intercalation rates is the excluded volume interactions that take place either in the bulk or on the surface of the system. For example, in a lattice-gas model (solid solution) the diffusional chemical potential of the intercalated ions is described by the following equation~\cite{bai2011,bazant2013} \begin{equation}\label{eq:sold_solution}
    \mu_{Li} = \mu^{\Theta}_{Li} + k_BT\ln\frac{c}{1-c}=k_BT\ln c+\mu^{ex}_{Li}
\end{equation}
where $\mu^{ex}_{Li}=\mu^{\Theta}_{Li}+k_BT\ln\gamma_{Li}$, and $\gamma_{Li}=1/\left(1-c\right)$ that corresponds to the excluded volume effects. 

\begin{figure*}[!ht]
    \centering
    \includegraphics[width=1\textwidth]{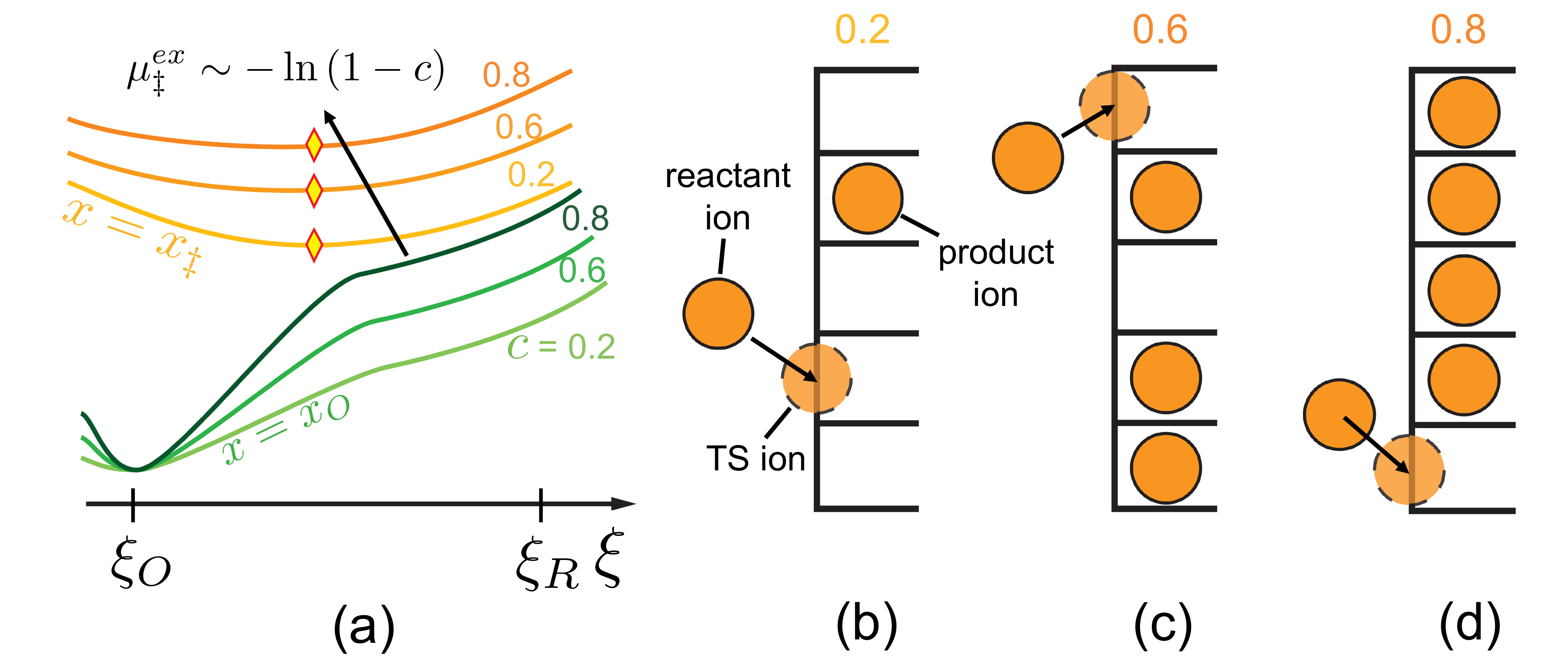}\hspace{8pt}
    \caption{(a) Schematic of the reactant and transition state energy landscape for only the ion transfer. With increasing products concentration $c_R\equiv c$, the entropic effects on the transition state increase the effective activation energy for the ion transfer process. As a result, the transition state excess chemical potential $\mu_{\ddag}^{ex}$ scales as $\ln(1-c)^{-1}$. (b)-(d) Physical picture of the entropic effects on the transition state. Three representative concentrations are shown. With increasing the concentration of products, the ions at the transition state start interacting entropically with their environment, decreasing the probability for the ion transfer to occur.}
    \label{fig:exclusion}
\end{figure*}

In the case of the transition state, the picture is similar to the bulk~\cite{bai2011,bazant2013,bazant2013,cogswell2012,lim2016origin,Li_nat_2018,Nadkarni2018,nadkarni_LCO2019,fraggedakis_stability2019}. More specifically, the idea of excluded sites on the transition state is demonstrated in figs.~\ref{fig:exclusion}, where we provide a combined energetic and physical picture. For simplicity, we consider the case of an isolated ion transfer to understand solely the excluded volume effects. Thus, fig.~\ref{fig:exclusion}(a) shows the energy landscape for both $x=x_O$ and $x=x_\ddag$ to demonstrate the effects of surface crowding on the reaction rate. We focus on three cases: 1) low, 2) intermediate, and 3) high concentration of intercalated ions $c_R\equiv c$. 

During ion transfer, $\xi=\xi_\ddag$, the transferring ion occupies a free site from the product state. At the same time, all the other sites are populated by the ions of the product state. For low $c$, it is clear that there is a high probability for the TS ions to find a free site to be transferred, fig.~\ref{fig:exclusion}(b) which corresponds to the light green curve in fig.~\ref{fig:exclusion}(a). Once the concentration of products increases, fig.~\ref{fig:exclusion}(c), the ions at the transition state start competing with those at the product state for free space. In other words, the entropic effects at the transition state decrease the probability of having a complete ion transfer by effectively increasing (medium green) the `activation' energy of the process, fig.~\ref{fig:exclusion}(a). At very high product concentrations, fig.~\ref{fig:exclusion}(d), it becomes very rare for the transition state ions to not be repelled back to their reactant state. This translates to even higher (darker green) transition state energies, fig.~\ref{fig:exclusion}(a). In analogy to the activity coefficient of the bulk chemical potential shown in eq.~\ref{eq:sold_solution}, and also as described in detail in fig.~\ref{fig:exclusion} and suggested in Ref.~\cite{bazant2013}, the excluded-volume effect during intercalation can be modeled with the following expression for the transition state activity coefficient
\begin{equation}
\gamma_{\ddag}=\left(1-c_R\right)^{-s}
\label{eq:tst_conc_eff}
\end{equation}
where $s$ is the number of sites the transition state ions occupy during insertion. The species concentration effects on the transition state theory lead to reaction-limited current that depends on species concentration, in agreement with experiments on ion intercalation materials~\cite{besnard2017multiscale}.

In summary, for ion intercalation in a solid, excluding one site in the transition state ($s=1$), with electrons provided by a metallic electrode source, the CIET current density, eq.~\ref{eq:R_tot}, can be approximated as~\cite{zeng2014simple} 
\begin{equation}
\begin{split}
\label{eq:aprx_intercalation}
i \simeq ek_0^*\sqrt{\pi\lambda} \left(1-c\right)
& \left(\frac{1}{1+e^{\eta_f}}
-
\frac{c}{1+e^{-\eta_f}}\right)\\
& \times{\mathrm{erfc}\left( \frac{{\lambda -\sqrt{\hat{\alpha}+\eta_f^2}}}{2\sqrt{\lambda}} \right )}
\end{split}
\end{equation}
where $c$ is the normalized concentration of Li ions in the material~\cite{bai2011,nadkarni_LCO2019,fraggedakis_stability2019,fraggedakis2020scaling}.

In the following sections, we test the predictions of coupled ion-electron transfer on describing Li ion intercalation in LFP on both the single particle and porous electrode scales. The Li intercalation reaction in LFP is modeled as
$$
    \mathrm{Li}^+_{sol} + \mathrm{Fe}^{+3}_\mathrm{s} + e^- \leftrightarrows \mathrm{Li}^+_{\mathrm{s}} + \mathrm{Fe}^{+2}_\mathrm{s}
$$
where $\mathrm{Li}^+_{sol/\mathrm{s}}$ denote the lithium ions in the electrolyte phase and in the particle respectively, while $\mathrm{Fe}^{+3/+2}_\mathrm{s}$ are the oxidized and reduced states of $\mathrm{Fe}$ in the crystal. The electrons originate from the carbon film that surrounds the  LFP particles~\cite{bai2014}. Finally, the thermodynamic model for LFP can be found in~\cite{bai2011,Nadkarni2018,cogswell2011}.

\subsection{Exchange current density}
Most BV-based reaction models admit the factorization of the form~\cite{bazant2013,bazant2017thermodynamic}
$$
  i=eF\left(c_O,c_R\right)G\left(\eta\right)
$$
where $F\left(c_O,c_R\right)$ is a function only of the system species concentrations, which is directly related to the exchange current density $i_0$, and $G\left(\eta\right)$ is solely a function of the applied overpotential~\cite{bazant2017thermodynamic,doyle1993}. When the CIET model is cast in the same form as BV, eq.~\ref{eq:itot_bv}, this factorization is not possible. More specifically, the charge transfer coefficient $\alpha$ becomes a function of both $c_O$ and $c_R$, leading to $G\left(c_O,c_R,\eta\right)$. In fact, this non-linear coupling between $\alpha$ and $\eta$ is one of the main reasons for distinguishing the models developed based on the quantum-mechanical picture of electron transfer from those which are purely phenomenological.

Table~\ref{tbl:models} summarizes several models that are commonly used in electrochemical ion intercalation. The first three are based on the BV formulation~\cite{doyle1993,bazant2013,lim2016origin}, where only $F\left(c\right)$ differs and $\alpha$ is a material parameter, while the last two correspond to ET kinetics, where $\alpha$ is now a function of the intercalated lithium concentration. 

\begin{table*}
\centering
\begin{tabular}{lll}
\hline
\multicolumn{1}{c}{Model}                                             & \multicolumn{1}{c}{} & Ref. \\
\multicolumn{1}{c}{$F\left(c\right)$}                                 & \multicolumn{1}{c}{$G\left(c,\eta\right)$} & \\ \hline
$k^*_{BV,0}\left(1-c\right)e^{\mu/2}$ & $\,\,\,\,\,\,\,\,\,\,\,\,\,\,\,\,\,\,\,\,\,\,\,\,\exp\left(-\alpha \eta\right)-\exp\left(\left(1-\alpha\right)\eta\right)$ &~\cite{bai2011,bazant2013}  \\
$k^*_{BV,0}\sqrt{c\left(1-c\right)}$                            & \multicolumn{1}{c}{"\,\,\,\,\,\,\,\,\,\,\,\,"} &~\cite{doyle1993}           \\
$3k^*_{BV,0}\left(1-c\right)\sqrt{c\left(1-c\right)}$           & \multicolumn{1}{c}{"\,\,\,\,\,\,\,\,\,\,\,\,"} &~\cite{lim2016origin}       \\
$k^*_0\frac{\left(1-c\right)e^{\mu/2}}{\sqrt{c}}$               & $\exp{\left({\frac{{\eta^2}}{4 \lambda}}\right)}
    \left[ \exp{\left( -\alpha\left(c\right)\eta \right)} - \exp{\left( (1-\alpha\left(c\right))\eta \right)}\right]$ &~\cite{smith2017multiphase} \\ 
$k^*_0\left(1-c\right)$                                         &  \multicolumn{1}{c}{"\,\,\,\,\,\,\,\,\,\,\,\,"} & eq.~\ref{eq:itot_bv},~\cite{smith2017multiphase}
\end{tabular}
\caption{Constitutive relations for ion intercalation. For clarity, all BV-based and CIET-based models are described with constant reaction constants, $k_{BV,0}^*$ and $k_0^*$, respectively. The third model in the table, which is a variant of ET kinetics, is one of the versions suggested in~\cite{smith2017multiphase} and used for the simulations therein. }
\label{tbl:models}
\end{table*}

When the chemical potential $\mu$ is modeled using regular solution theory~\cite{bai2011,cogswell2012,cogswell2013}, the first two models differ only by the factor which describes the Li-Li interactions. The regular solution term is responsible for the existence of spinodal points in a thermodynamics system~\cite{cahn1958}. The exchange current density $i_0$ used by Srinivasan \& Newman~\cite{doyle1993,srinivasan2004}, corresponds to an ideal lattice gas model with excluded volume effects. By postulating $G\left(\eta\right)$ to be similar to the BV model, Lim et al.~\cite{lim2016origin} mapped $F\left(c\right)$ using the experimentally extracted current. 

Both the first~\cite{bazant2013} and the third~\cite{lim2016origin} models capture the auto-inhibition mechanism which is responsible for the suppression of phase separation under non-equilibrium conditions~\cite{bazant2017thermodynamic}. The former, though, overestimates the range of $c$ for which the reaction rate decreases with concentration~\cite{lim2016origin}. Related to the differences between the ET models, the latter model introduced in Ref.~\cite{smith2017multiphase} starts with the same general expression predicted by CIET theory, Eq. (\ref{eq:approximation}), but is modified to replicate the exchange current density of the BV model~\cite{bazant2013} by eliminating the rough $\sqrt{c}$ dependence of $G\left(c,\eta\right)$, which is only true for large $\lambda$.  It is thus interesting to test how this {\it ad hoc} departure from the CIET theory affects predictions of experimental data in this section and the next.

\begin{figure}[!ht]
    \centering
    \hspace{0.095in}\includegraphics[width=3.5in]{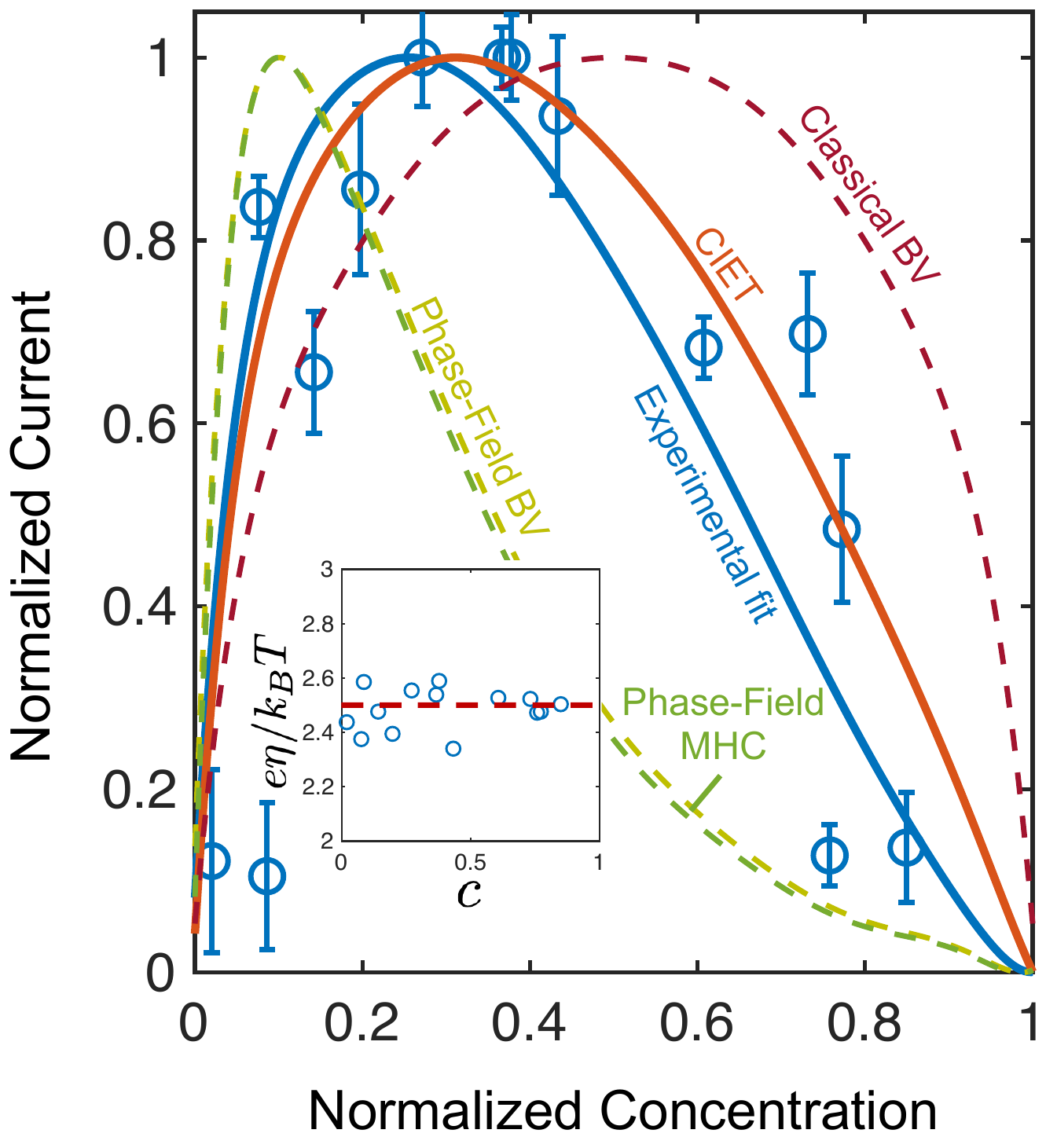}
    \caption{Comparison of the observed normalized current~\cite{lim2016origin} versus the predicted ones by the models in table.~\ref{tbl:models}. Solid lines describe the fitting on the experiments~\cite{lim2016origin} and the theoretical predictions of the models derived in section~\ref{sec:model}. With dashed lines are shown the predictions of the phenomenological models as introduced in~\cite{bazant2013,doyle1993,smith2017multiphase}. For clarity, we use the following abbreviation for the models compared here: 1) Experiment~\cite{lim2016origin}, 2) Phase-Field BV~\cite{bai2011,bazant2013}, 3) Classical BV~\cite{srinivasan2004,doyle1993,newman2004}, 4) Phase-Field MHC~\cite{smith2017multiphase}. The model of eq.~\ref{eq:aprx_intercalation} is not fitted to the experimental data. The parameters for the thermodynamics model of LiFePO$_4$ can be found in~\cite{bai2011}, while the used value for the reorganization energy is taken from~\cite{bai2014}, and is $\lambda\simeq8.3k_BT$. The inset figure corresponds to the normalized experimentally measured overpotential $e\eta/k_BT$ as a function of the local Li ion concentration $c$~\cite{lim2016origin}, where the dashed dark red line corresponds to the average value of $e\eta/k_BT\simeq 2.5$. }
    \label{fig:curr_conc}
\end{figure}

Fig.~\ref{fig:curr_conc} shows the predictions of the normalized current $i/i_{max}$ as a function of $c$. The maximum current $i_{max}$ is defined at the normalized concentration $c$ where $i$ attains its maximum value. We compare the model predictions with the experimentally measured data of the current density for LFP~\cite{lim2016origin} (blue dots). The measured local concentration $c$ and overpotential $\eta$ are coupled, as the discharging was performed under constant current. Thus, fig.~\ref{fig:curr_conc} is constructed by directly using the experimentally observed values of $\left(c,\eta\right)$ into the models presented in table~\ref{tbl:models}. The solid lines illustrate the fitting (blue) on the experimental data as performed in~\cite{lim2016origin} along with the theoretical predictions of the CIET model (orange). The dashed lines show the predictions of the other models of table~\ref{tbl:models}.

It is apparent that the model of our work along with the empirical fit~\cite{lim2016origin} can capture the correct behavior of the normalized current vs. $c$. It noteworthy that the developed model has not been calibrated to the experimental data and the predictions are based on the material parameters found in Refs.~\cite{cogswell2012,bai2014}. The reorganization energy of LFP is equal to $\lambda = 8.3k_BT$~\cite{bai2014}. Regarding the predictions of the other three models, they either overestimate or underestimate the concentration $c_{max}$ at which the current is maximized. 

The quantitative agreement of our model with the experimental data highlights the importance of considering electron transfer coupled with ion transfer for ion intercalation in solids. This aspect is missing in earlier models of ion intercalation based on BV kinetics~\cite{doyle1993,fuller1994relaxation,srinivasan2004,bazant2013,bai2011}, which we show here could arise in certain limits of the CIET theory for either large reorganization energy or large ion transfer energy at moderate overpotentials.   Several similar, thermodynamically consistent generalizations of Marcus~\cite{bazant2013} and MHC \cite{smith2017multiphase} kinetics have also been postulated for ET-limited ion intercalation, but here we provide the first general microscopic theory capable of describing all of these limits and predicting the proper form of the rate expression.  In the next section, we provide further quantitative support for the CIET theory by resolving the controversy over the original measurements revealing MHC kinetics in Li-ion batteries~\cite{bai2014}.

\subsection{Chronoamperometry with porous electrodes}
Porous electrode theory~\cite{newman1975,ferguson2012,smith2017multiphase} is widely used for predicting the behavior of macroscopic quantities (e.g. current/voltage response, and (dis)charging capacity) which are important in energy-related applications (e.g. Li-ion batteries~\cite{Nitta2015}). In general, the resulting voltage and current follow complex dynamics which are a result of coupled processes across multiple scales. In Li-ion batteries for example, the cathode and anode consist of multiple particles where Li insertion and solid diffusion are important. The primary particles and secondary agglomerates are connected to each other through the electrolyte, as well as conducting additives. Here, our goal is to demonstrate how CIET theory performs on the porous electrode scale. More specifically we compare the predictions of the model with the chronoamperometry experiments of~\cite{bai2014} on LFP.

\begin{figure*}[!ht]
    \centering
    \hspace{0.095in}\includegraphics[width=\textwidth]{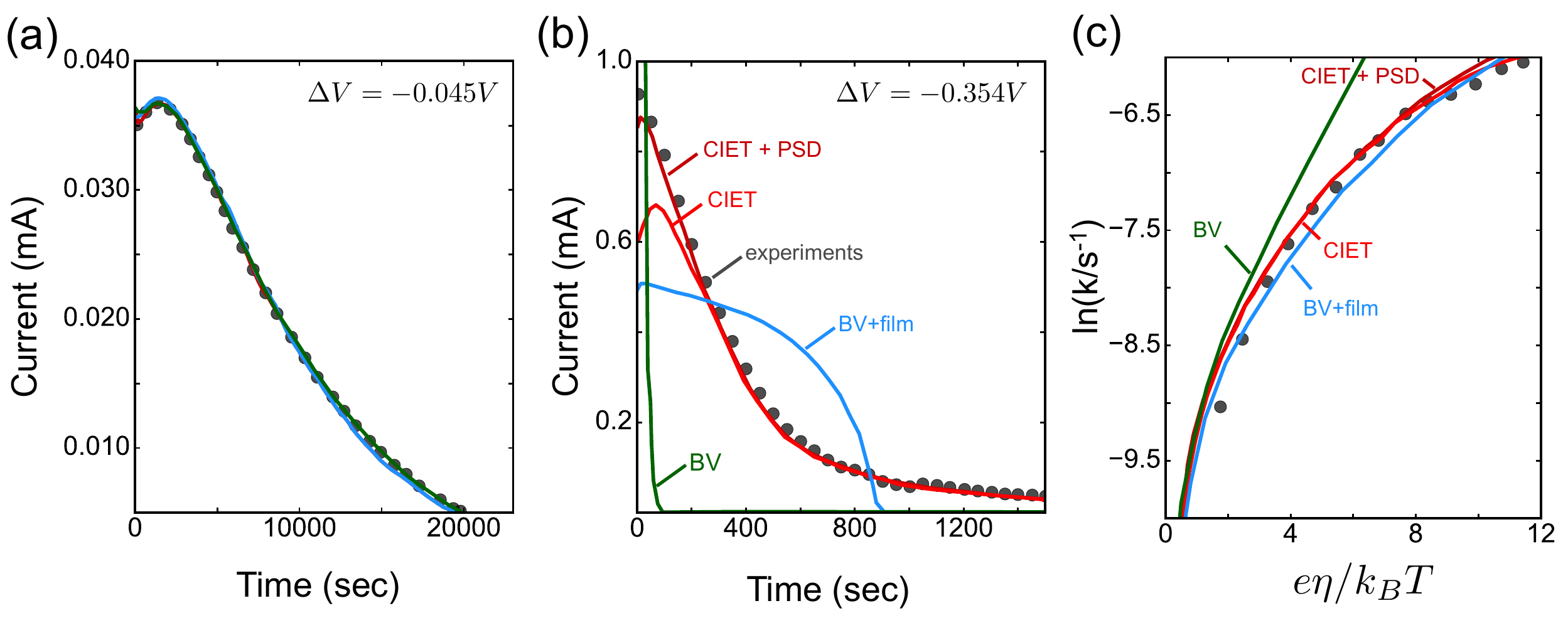}
    \caption{Comparison of reaction models in chronoamperometry experiments and simulations. Experimental data are from ref.~\cite{bai2014}. (a)-(b) Transient simulated (solid lines) and experimental (black circles) current responses, under a voltage step of $\Delta V = -0.045$ V and $\Delta V = -0.354$ V, respectively, for the CIET (red line), CIET with particle size distribution (PSD) (dark red line), BV+film (blue line) and BV (green line) reaction models. (c) Tafel plot constructed using the method presented in~\cite{bai2013,bai2014} to extract the representative reaction rate constant $k$.}
    \label{fig:bv_mhc}
\end{figure*}

The system we are interested in contains $N$ LFP particles and is initially prepared at $V_{cell} = 3.422$ V and room temperature. At $t=0$, we apply a voltage step of magnitude $\Delta V$. We use the same voltage step values as those in~\cite{bai2014}. The voltage range under which the experiment is performed covers a large spectrum of Li concentrations inside the active material of the cathode. This is seen from the voltage-capacity curves of LFP, where for the largest voltage drop $V_{cell}+\Delta V_{max}=3.069$V, the final capacity of the intercalated Li in the cell lies in the spinodal~\cite{dreyer2010,bai2011,lim2016origin}.

As in the case of single particles, we are interested in comparing the predictions of different reaction models that are commonly used to describe ion intercalation kinetics. In particular, we compare the developed model against BV~\cite{bai2011,bazant2013} and BV with film resistance $R_f$ (BV+film)~\cite{doyle1996,ferguson2012,smith2017multiphase,heubner2015investigation}. The mathematical expression of the latter can be found in eqs. 35, 36, and 37 of Ref.~\cite{smith2017multiphase}. BV+film is known to reproduce curved Tafel plots~\cite{heubner2015investigation}, similar to the ones predicted by electron transfer limitations~\cite{bai2014}. Therefore, we test the models not only in terms of their capability to predict the Tafel plots, but also on the time evolution of the resulting current after we apply the voltage step. 

We perform the simulation using porous electrode theory, pioneered by Newman~\cite{newman1962,doyle1993}, as recently modified to describe phase separating materials ~\cite{ferguson2012,smith2017multiphase}. We refer the readers interested in the porous electrode theory to Refs.~\cite{newman2004,ferguson2012,smith2017multiphase} and for the numerical methods for discretizing the equations to~\cite{fraggedakis2015flow,fraggedakis2017discretization} for more details. The cell has a diameter of $1.27$ cm, while each electrode has thickness approximately around $L_{\text{electrode}}=4$ $\mu$m. The size of the LFP primary particles is described by a log-normal distribution~\cite{crow1987lognormal} with average diameter $\left<d\right>=1$ $\mu$m, and variance of $\sigma_d^2 = 250$nm, a value found by fitting the theory to the experiments. The total number of particles used in the simulations is of the order $N\sim O\left(10^4\right)$. The simulation results using the particle size distribution are denoted with PSD, otherwise we set $\sigma_D^2=0$. All the parameters related to the geometry of the cell, electrode and the particle size are reported in the Methods section of~\cite{bai2014}. In addition to the variance of the particle population, we also adjust the constant reaction rate prefactor $k_0^*$ to fit the experiments, and we find it to be $k_0^* = 8\times10^{-3}$A/m$^2$. The coupled ion-electron transfer and the BV models were calibrated on half of the available experimental data sets~\cite{bai2014} that correspond to the lower values of the applied $\Delta V$, while the BV+film model was calibrated on the Tafel plot of~\cite{bai2014}. The fitting was performed using a common non-linear least squares procedure~\cite{nocedal2006numerical}. This fitted value of $k_0^*$ is very close to the one calculated in Ref.~\cite{lim2016origin} by assuming a reaction-limited process. For the Butler-Volmer with film resistance, we use $R_f=7$ $\Omega$ m$^2$ to fit the Tafel plot. 

Additionally, we neglect diffusion limitations in the electrolyte within the electrode, and we set the number of volumes in the porous electrode model equal to one. This assumption is based on the estimate of the liquid diffusion timescale for the electrolyte within the electrode $\tau_{\text{electrolyte}}=L_{\text{electrode}}^2/D_{Li^+}$, where $D_{Li^+}$ is the liquid electrolyte diffusion coefficient of the solvated Li ions. Using $D_{Li^+}\simeq 5\times10^{-11}$ m$^2$/s as a characteristic value, we find $\tau_{\text{electrolyte}}\simeq0.3$ s, which is much shorter than the operation timescale of the cell (around $1200$ seconds for the largest applied overpotential). Under these conditions we can safely assume nearly uniform Li concentration across the electrolyte phase. 

In general, porous electrode experiments involve thousands of primary particles of variable sizes. As discussed in~\cite{bai2013}, the resulting macroscopic quantities of the cell are affected by several factors such as particle activation/nucleation, phase-separation, variable particle size, inhomogeneous SEI formation amongst different particles, etc. Therefore, the usage of a statistical method is important in order to extract the true reaction constants of the studied reaction. We applied the same protocol to our simulation results to construct in a similar fashion the Tafel plot shown in fig~\ref{fig:bv_mhc}(c). More specifically, after performing the simulations, we extract the resulting current, and we fit it with eq.(4) of~\cite{bai2014}. Then, we use the fitted $k$ values (reaction constant in the population model) to construct the Tafel plot.

Fig.~\ref{fig:bv_mhc}(a) illustrates the current $I$ vs. time $t$ after we apply a step of $\Delta V = -0.045V$ ($e\eta ~\sim 2k_BT$). The experimental measurements are shown with black circles, while the model predictions are shown with continuous lines. As shown in fig.~\ref{fig:bv_mhc}(a), for conditions near equilibrium (small overpotential) all models behave similarly, reproducing the experimentally observed current-time response. At large applied overpotential, however, the predictions of each model start to deviate from each other. 

Fig.~\ref{fig:bv_mhc}(b) shows the transients of the current under $\Delta V_{max}=-0.354$ V, a value almost 10 times larger than the previously discussed voltage step. Under these conditions, the predictability of the developed coupled ion-electron transfer is apparent (red line), where it is able to reproduce the experimentally observed current for the largest percentage of the studied time interval. Regarding the predictions of BV kinetics (green line), we know that the model predicts exponentially increasing current with increasing overpotential. Under the experimental conditions the current is overestimated for the applied overpotential, and decays rapidly at very early times ($t\sim100$ s), fig.~\ref{fig:bv_mhc}(b). Finally, the predictions on current vs. $t$ using BV+film are shown with the blue line. The predicted values of current are not able to capture the experimentally measured values, making clear the discrepancies that can be introduced by using only Tafel measurements to characterize the mechanism of a reaction.

This last comparison raises questions on the interpretation of experimental data by using classical methods such as Tafel analysis. Fig.~\ref{fig:bv_mhc}(c) shows the constructed Tafel plot by using the statistical method introduced in~\cite{bai2013} and used in~\cite{bai2014}. Again, the red line represents the predictions using CIET and the blue line those of BV+film. The agreement between CIET and BV+film is consistent with the discussion in Ref.~\cite{heubner2015investigation}, where it was shown that curved Tafel plot data can be fitted by Butler-Volmer with film resistances included. However, as shown in fig.~\ref{fig:bv_mhc}(b), the latter model is not able to capture the experimentally observed trends of the current transients under large values of overpotential ($e\eta/k_BT\gg1$). In both current-time tests, as well as on comparing the extracted Tafel plots, we find that CIET predictions are consistent with the experimentally observed behavior on the porous electrode scale.

As a final remark on the porous electrode analysis, we would like to raise general questions related to the characterization of the reaction kinetics by classical electro-analytical methods~\cite{bard2001}. It is common practice to use either electrochemical impedance measurements, Tafel analysis, or cyclic voltammetry to characterize the processes present in an electrochemical system. Here, we show that in a reaction-limited system~\cite{lim2016origin}, the Tafel analysis and EIS~\cite{heubner2015investigation} alone are not able to resolve the rate-determining step of Li-ion intercalation and predict other types of measurements. In a complicated system such as a porous electrode, it is difficult to use the classical electroanalytical machinery to deconvolute different processes that take place across multiple scales involving highly nonlinear couplings of reaction and diffusion. 

\section{Discussion}
\label{sec:discussion}
The application of CIET theory is by no means limited to lithium intercalation in LFP~\cite{bazant2013,smith2017intercalation,Nadkarni2018}. The generality of the reaction rate expressions presented in Sec.\ref{sec:model} makes the theory applicable to CIET reactions in both concentrated solids and liquids. As described earlier, candidate processes in which coupled ion-electron transfer might be the rate-determining step include: (i) lithium intercalation in other host materials, such Li$_x$CoO$_2$ (LCO)~\cite{nadkarni_LCO2019}, Li$_x$C$_6$ (graphite)~\cite{ferguson2014,smith2017intercalation,thomas-alyea2017insitu}, Li$_x$TiO$_2$ (anatase)~\cite{de2017explaining}, and Li$_{4+3x}$Ti$_5$O$_{12}$ (LTO)~\cite{Li2018,vasileiadis2018toward,gonzalez2020lithium}, used in Li-ion batteries, as well as in neuromorphic computing devices~\cite{fuller2017li,nadkarni_LCO2019,gonzalez2020lithium,fraggedakis2020dielectric}, (ii) sodium intercalation in Na-ion batteries~\cite{palomares2012ion,zhang2018sodium} or capacitive deionization~\cite{porada2017nickel,smith2017theoretical,singh2018theory},(iii) multivalent aluminum ion intercalation in Al-ion batteries~\cite{jayaprakash2011rechargeable,leisegang2019aluminum} (iv) oxygen insertion in perovskite oxides used in fuel cells~\cite{Hwang2017,Ormerod2003}, or oxygen reduction using perovskites as catalysts for metal-air batteries~\cite{Suntivich2011}.

The material parameters which enter the model are directly connected with the microscopic nature of the species which participate in the reaction. In particular, via the explicit usage of chemical potentials, the non-ideal nature of the species is included. Also, the electron energy levels are taken into account by describing the band structure of the donor of the electrons via the density of states of the material. Finally, the interactions between the electrons with their environment is described through the reorganization energy. All this microscopic information establishes coupled ion-electron transfer as a quantitative, physics-based model for intercalation reaction kinetics. 

There are several Li ion intercalation studies where coating the intercalation material with anionic additives increases the rate performance~\cite{Li2011_coating,park2012enhanced,Song2016,Chi2014,Wang2012,Lin2013,wang2015olivine,Goodenough2013}. Until now, existing models cannot explain the reason why this occurs. More specifically, the parameters in BV-based models cannot be directly related to the physical details of the reaction event process, while ET models describe only the electron transfer event without considering the fate of the ions. On the other hand, the idea of coupled ion-electron transfer takes into account the microscopic physics of both the ion and electron transfer, as we consider both processes to occur simultaneously. Through CIET, we are able to give a possible explanation for why the anionic-coating rate enhancement occurs. The model includes the energies $w_{O/R}$ that correspond to the `adsorption' of the ion at the reaction interface ($\xi=\xi_O$). 

In general, $w_{O/R}$ corresponds to the repulsive/attractive interactions between the solvated ions and interface atoms, and the diffuse double-layer effects on the ions that participate in the reaction. For Li-ion intercalation, when anionic groups, for example $N-$ or $S-$ groups~\cite{park2012enhanced}, are added on the surface of the intercalation material, the energy barrier $w_{O/R}$ decreases, leading to an increase of the effective reaction rate constant. This behavior is in qualitative agreement with both experimental and \textit{ab-initio} studies on Li intercalation in LiFePO$_4$~\cite{park2012enhanced,Li2011_coating,wang2015olivine}. 

Coupled ion-electron transfer can be used to provide insights on the design and engineering of interfaces where electrochemical reactions take place. Very recently, it has been shown that by understanding the functional form of the reaction rate expressions, one is able to control, and consequently engineer, the physics of interfaces where reactions take place~\cite{bazant2017thermodynamic}. Representative examples are the lithiation of LFP and LiNi$_{1/3}$Mn$_{1/3}$Co$_{1/3}$O$_2$~\cite{tsai2018single,zhang2020revealing} particles, as well as the operation of Li-air batteries, where the thermodynamic stability of the system is controlled by varying the applied current~\cite{bai2011,lim2016origin,horstmann2013,fraggedakis_stability2019,zhao_population_2019}. In terms of CIET, by understanding the concentration dependencies of both the reorganization energy and the density of states of the electron donor, we will be able to control interface structure by inducing or suppressing phase separation/island formation. 

There are cases where increased interfacial anisotropy is desired. For example in electro-catalytic applications the interface structure of the active area affects the efficiency of processes like dealloying~\cite{Erlebacher2001} or light absorption~\cite{Mandal2017}. In other cases surface anisotropy can lead to mechanical failure, e.g. in all-solid-state Li-ion batteries where loss of contact between the active material and the solid electrolyte leads to irreversible capacity loss~\cite{Koerver2017}. Thus, the present theoretical framework of coupled ion-electron transfer draws a connection between the structural information of the species participating in the reaction with the operational conditions, providing direct ways to engineer surfaces using electrochemical methods~\cite{fraggedakis_stability2019}.

The idea of the coupled ion-electron transfer can be extended to describe the diffusion of ion-$e^-$ pairs in solids. This description can give important insights on the limitations of technologies such as solid-state batteries~\cite{Luntz2015}, where the electronic conductivity of solid electrolytes~\cite{Kornyshev1978,kornyshev1981conductivity} is not fully understood yet~\cite{Han2019}.

{
As a last remark, we would like to stress again that, while the physical picture of coupled ion-electron transfer is general, our mathematical formulation expressing the rates via  eq.~\ref{eq:R_total_final}(a) \& (b) corresponds to non-adiabatic electron transfer. There are important cases where the electronic states of the RedOx states are strongly coupled~\cite{santos2009model,lam2019theory}, as in specific ion adsorption, and the electron transfer occurs adiabatically. In such situations, the electron transfer event depends strongly on the electronic interactions between the RedOx species as well as with the solvent~\cite{SchmicklerText,Schmickler1996}. These effects have been previously discussed in the context of coupled proton-electron transfer for describing hydrogen evolution, where the proton transfer is characterized by its distance from the electrode and the electron transfer occurs adiabatically~\cite{huang2018interplay,lam2019theory}. In this case, the mathematical formulation of the forward and backward reaction rates is typically analyzed using model Hamiltonians, such as the Newns-Anderson model, combined with mean-field electrostatics for the double layer effects on the proton transfer~\cite{lin2016electrical}.
}

\section{Summary}
\label{sec:conclusion}

In this work, the theory of electron transfer has been extended to incorporate ion transfer effects on the reaction kinetics. In particular, by expanding the reaction space to include additional coordinates in addition to the polarization one, we include phenomena such as surface crowding, (de)solvation effects, misfit stress contributions, etc. on the transition state. Moreover, the thermodynamics of the species are incorporated in the reaction kinetics formalism, allowing for the description of phase separation and its effects on the reaction rate. The results presented here illustrate the importance of coupled ion-electron transfer kinetics in ion intercalation kinetics~\cite{bazant2013,lim2016origin,bai2014}. {The key expressions for the total reaction rate derived from our theory are 
$$
    i_{red,\varepsilon} = 
    \frac{ek^*_0c_O}{\gamma_{\ddag}}  n_e
    \exp\left( -\frac{{\left( \lambda + e\eta_f - x_\varepsilon \right)}^2}{4\lambda k_BT} \right)
$$
$$    
    i_{ox,\varepsilon} = 
    \frac{ek^*_0c_R}{\gamma_{\ddag}}  \left[1-n_e\right]
    \exp\left( -\frac{{\left( \lambda - e\eta_f + x_\varepsilon \right)}^2}{4\lambda k_BT} \right)
$$
$$
    i= \int_{-\infty}^{\infty}
    \left(i_{red,\varepsilon}-i_{ox,\varepsilon}\right)\rho \,\,d\varepsilon
$$
where $x_\varepsilon = \varepsilon-E_f$,  $e\eta_f=e\eta -  k_BT\ln\left(c_O/c_R\right)$, $e\eta= eV^\Theta + k_B T\ln\left(\frac{\gamma_R c_R}{\gamma_Oc_O}\right) + e\left(z_R\phi_R-z_O\phi_O\right)-E_f$, and $n_e=\left(1+e^{\varepsilon-E_f}\right)^{-1}$.} For practical purposes, the reaction rate prefactor $k^*_0$ can be fitted to experiments or predicted from first-principles as described in eq.~\ref{eq:w_or}. Additionally, the transition state activity coefficient can take into account the effects of the environment on the ion transfer event. One example is the exclusion of a free site during the transfer of an ion to its product state, where $\gamma_\ddag$ scales with the number of available free sites, $\gamma_\ddag\propto \left(1-c_R\right)^{-1}$. For insertion of ions in solids, pre-existing strains developed due to concentration fluctuations can also affect the transition state barrier.

The usage of the theory is demonstrated by modeling the insertion of ions in solid materials, a process present in several applications of technological importance. By comparing the predicted current density to available experimental data~\cite{lim2016origin,bai2014} on lithium intercalation in primary FePO$_4$ particles, we demonstrated the capability of CIET to accurately describe, on microscopic (single particle) and macroscopic (porous electrode) levels, the (dis)charging process of Li-ion batteries. In particular, the model predicts the experimentally observed normalized total current without using any adjustable parameters. Additionally, the surface crowding effects upon Li insertion were found to be crucial in correctly predicting the auto-inhibitory nature of the phenomenon.

{\section*{Appendix}
\subsection*{Derivation of De Donder relation for coupled ion-electron transfer kinetics}

An essential constraint on any thermodynamically consistent model of reaction kinetics is that the forward and backward rates satisfy the de Donder relation expressing microscopic reversibility~\cite{sekimoto2010}, which takes the following form for an electrochemical reaction~\cite{bazant2013},
\begin{equation}
\frac{R_{red}}{R_{ox}}= e^{-\frac{e\eta}{k_BT}}
\end{equation}
In order to prove De Donder relation for the ratio of $R_{red/ox} = \int_{-\infty}^{\infty} R_{red/ox,\varepsilon}\rho\,\,d\varepsilon$, it is useful to group all $\varepsilon$-dependencies together. In particular, we can recast eqs.~\ref{eq:R_total_final} in the following form
\begin{widetext}
\begin{subequations}
\label{eq:R_total_final_dedonder}
\begin{equation}\label{eq:R_fwd_dd}
\begin{split}
R_{red,\varepsilon} = &
    \frac{\tilde{k}_0 e^{-\alpha_\xi\Delta E_{IT}/k_BT}}{\gamma_{\ddag}}
    \frac{c_O}{1+e^{x_\varepsilon/k_BT}}
    e^{ -\frac{{\left( \lambda + e\eta_f \right)}^2}{4\lambda k_{B}T} }
    e^{-\frac{{x_\varepsilon \left( 2\left(\lambda+e\eta_f\right) - x_\varepsilon \right)}}{4\lambda k_{B}T} }
\end{split}
\end{equation}
\begin{equation}\label{eq:R_bck_dd}
\begin{split}
R_{ox,\varepsilon} = 
    \frac{\tilde{k}_0 e^{-\alpha_\xi\Delta E_{IT}/k_BT}}{\gamma_{\ddag}} 
    \frac{c_R}{1+e^{x_\varepsilon/k_BT}}
    e^{ -\frac{{\left( \lambda - e\eta_f \right)}^2}{4\lambda k_{B}T} }
    e^{-\frac{{x_\varepsilon \left( 2\left(\lambda+e\eta_f\right) - x_\varepsilon \right)}}{4\lambda k_{B}T} }
\end{split}
\end{equation}
\end{subequations}
\end{widetext}
where we used the definition of $n_e=1/(1+e^{x_\varepsilon/k_BT})$. Then, the ratio $R_{red}/R_{ox}$ can be simplified as follows:
\begin{widetext}
\begin{equation}
\begin{split}
    \frac{R_{red}}{R_{ox}}&=\frac
    {\int_{-\infty}^\infty \frac{\tilde{k}_0 e^{-\alpha_\xi\Delta E_{IT}/k_BT}}{\gamma_{\ddag}} 
    \frac{c_O}{1+e^{x_\varepsilon/k_BT}} e^{ -\frac{{\left( \lambda + e\eta_f \right)}^2}{4\lambda k_{B}T} }
    e^{-\frac{{x_\varepsilon \left( 2\left(\lambda+e\eta_f\right) - x_\varepsilon \right)}}{4\lambda k_{B}T} } \rho\,\,d\varepsilon}
    {\int_{-\infty}^\infty  \frac{\tilde{k}_0 e^{-\alpha_\xi\Delta E_{IT}/k_BT}}{\gamma_{\ddag}} 
    \frac{c_R}{1+e^{x_\varepsilon/k_BT}}
    e^{ -\frac{{\left( \lambda - e\eta_f \right)}^2}{4\lambda k_{B}T} }
    e^{-\frac{{x_\varepsilon \left( 2\left(\lambda+e\eta_f\right) - x_\varepsilon \right)}}{4\lambda k_{B}T} } \rho\,\,d\varepsilon} \\
    & = \frac
    {c_O e^{ -\frac{{\left( \lambda + e\eta_f \right)}^2}{4\lambda k_{B}T} } \int_{-\infty}^\infty  
    \frac{1}{1+e^{x_\varepsilon/k_BT}} 
    e^{-\frac{{x_\varepsilon \left( 2\left(\lambda+e\eta_f\right) - x_\varepsilon \right)}}{4\lambda k_{B}T} } \rho\,\,d\varepsilon}
    {c_R e^{ -\frac{{\left( \lambda - e\eta_f \right)}^2}{4\lambda k_{B}T} } \int_{-\infty}^\infty  
    \frac{1}{1+e^{x_\varepsilon/k_BT}}
    e^{-\frac{{x_\varepsilon \left( 2\left(\lambda+e\eta_f\right) - x_\varepsilon \right)}}{4\lambda k_{B}T} } \rho\,\,d\varepsilon} = 
    \frac
    {c_O e^{ -\frac{{\left( \lambda + e\eta_f \right)}^2}{4\lambda k_{B}T} } }
    {c_R e^{ -\frac{{\left( \lambda - e\eta_f \right)}^2}{4\lambda k_{B}T} } }\\
    & = \frac{c_O}{c_R}e^{-\frac{e\eta_f}{k_BT}} = \frac{c_O}{c_R}e^{-\frac{e\eta}{k_BT} + \ln\frac{c_R}{c_O}}=e^{-\frac{e\eta}{k_BT}}
\end{split}
\end{equation}
\end{widetext}
}
which is the De Donder relation.

\section*{Contributions}
D.F. and M.Z.B. formulated the framework of coupled ion-electron transfer, building on earlier attempts by R.B.S., Y.K. and P.B.  D.F. and M.M. integrated the quantum mechanical details. D.F., W.C.C and M.Z.B. justified the theory for ion intercalation materials, and D.F. performed the LFP simulations, building on the those of R.B.S. {D.F., Y.S.-H. and M.Z.B. connected the developed framework with the notation present in classical electrochemistry literature.} Y.Z. and Y.S.-H. provided helpful comments on the narrative and mechanisms. D.F. wrote the manuscript, and M.Z.B. supervised the study and revised the text. All authors contributed to the final manuscript.

\section*{Conflicts of interest}
There are no conflicts to declare.

\section*{Acknowledgments}
The research was supported by the D$_3$BATT program of the Toyota Research Institute and by the Shell International Exploration \& Production, Inc. The authors would like to thank Yiyang Li and Dean (Haitao) Deng for providing the raw data from the STXM experiments, Neel Nadkarni, Tao Gao, Tingtao Zhou, and Ryan M. Stephens for discussions related with the validity and application of the theory. 

\newpage

\bibliographystyle{elsarticle-num}
\bibliography{sources}
\end{document}